\documentclass[
 amsmath,amssymb,
 reprint,
 superscriptaddress,aps,
 nofootinbib,prd%
]{revtex4-2}

\usepackage{rotating} 
\usepackage{array}
\usepackage{graphicx}
\usepackage{booktabs}
\usepackage{amsmath}
\usepackage{arydshln}
\usepackage{latexsym}
\usepackage{mathrsfs}
\usepackage{amssymb}
\usepackage{multirow}                   
\usepackage{pifont}                     
\usepackage{dcolumn}
\usepackage{bm}
\usepackage{hyperref}

\usepackage[T1]{fontenc}
\usepackage[dvipsnames,table]{xcolor}   
\usepackage{youngtab}
\usepackage{comment}
\usepackage[normalem]{ulem}
\usepackage{longtable}
\usepackage{tikz}
\usepackage[compat=1.1.0]{tikz-feynman}
\usepackage[compat=1.1.0]{tikz-feynhand}

\definecolor{red}{rgb}{0.6,.0706,.1373}
\definecolor{blue}{rgb}{0,0.396,0.741}
\definecolor{PadovaRed}{rgb}{0.60546875,0,0.078125}

\colorlet{mylinkcolor}{violet}
\colorlet{mycitecolor}{violet}
\colorlet{myurlcolor}{violet}

\hypersetup{
  linkcolor  = PadovaRed,
  citecolor  = PadovaRed,
  urlcolor   = PadovaRed,
  colorlinks = true
}
\pretocmd{\tableofcontents}{\hypersetup{linkcolor=black}}{}{}
\apptocmd{\tableofcontents}{\hypersetup{linkcolor=red}}{}{}

\newcommand{\mc}[1]{\mathcal{#1}}
\newcommand{\cL}{\mathcal{L}}
\newcommand{\cO}{\mathcal{O}}

\newcommand{\cC}{{\mathcal C}}
\newcommand{\cS}{{\mathcal S}}

\newcommand{\gev}{\mathrm{GeV}}
\newcommand{\tev}{\mathrm{TeV}}
\newcommand{\E}[1]{\times 10^{#1}}
\newcommand{\hermc}{\text{h.c.}}
\newcommand{\eminus}{\vcenter{\hbox{\scalebox{0.6}[1]{$ - $}}}}	
\newcommand{\rep}[1]{\mathbf{#1}}
\newcommand{\repbar}[1]{\overline{\mathbf{#1}}}
\newcommand{\sscript}[1]{{\scriptscriptstyle \mathrm{#1}}}

\newcommand{\SU}{\mathrm{SU}}

\newcommand{\sig}{\sigma}
\newcommand{\lzm}{\left(}
\newcommand{\dzm}{\right)}
\newcommand{\lzs}{\left[}
\newcommand{\dzs}{\right]}

\newcommand{\re}{\mathrm{Re}}
\newcommand{\im}{\mathrm{Im}}

\begin{document}

\title{UV origins of CP-violating leptonic Yukawa couplings}

\author{Nejc Ko\v snik}%
 \email{nejc.kosnik@ijs.si}
 \affiliation{%
  Jo\v zef Stefan Institute, Jamova 39, 1000 Ljubljana, Slovenia
}%
\affiliation{%
  Faculty of Mathematics and Physics, University of Ljubljana, Jadranska 19, 1000 Ljubljana,
Slovenia
}%
\author{Ajdin Palavri\'c}%
\email{ajdin.palavric@unibas.ch}
\affiliation{%
Department of Physics, University of Basel, Klingelbergstrasse 82, CH 4056 Basel, Switzerland
}%
\author{Aleks Smolkovi\v c}%
\email{aleks.smolkovic@ijs.si}
\affiliation{%
Jo\v zef Stefan Institute, Jamova 39, 1000 Ljubljana, Slovenia
}%


\begin{abstract}
Probing the CP nature of the Higgs Yukawa couplings is a promising avenue for unraveling new physics effects. In this work, we investigate the possible single- and two-field UV origins of CP-violating leptonic Yukawa couplings, using the Standard Model Effective Field Theory as a stepping stone. We demonstrate a rich set of constraints on the UV model parameters, including direct Higgs measurements, electric and magnetic dipole moments of leptons, charged lepton flavor violating observables, and electroweak precision tests. Studying representative flavor assumptions we find that the precision constraints are often, but not always, more constraining than the dedicated LHC analyses of modified leptonic Yukawa couplings.
\end{abstract}

\maketitle
\section{Introduction}
\label{sec:intro}
The discovery of the Higgs boson at the LHC~\cite{ATLAS:2012yve,CMS:2012qbp, CMS:2013btf} represents one of the most important scientific breakthroughs in recent decades. Many of its properties have since been extensively studied and remain consistent with the predictions of the Standard Model (SM)~\cite{ATLAS:2015yey, ATLAS:2016neq, CMS:2018uag,ATLAS:2019nkf,ATLAS:2022vkf,LHCHiggsCrossSectionWorkingGroup:2016ypw}. Nevertheless, there is still ample room for manifestations of New Physics (NP), for instance through deviations in the Yukawa couplings of the light fermion generations~\cite{Erdelyi:2024sls,Erdelyi:2025axy}, or via CP-violating interactions, which are absent in the SM. Of particular interest are CP-violating modifications of the Yukawa couplings, which have attracted considerable experimental~\cite{ CMS:2020cga,CMS:2021sdq,ATLAS:2022akr,ATLAS:2023cbt} and theoretical~\cite{Kramer:1993jn,Grzadkowski:1995rx,Berge:2011ij,Berge:2014sra, Berge:2014wta,Ellis:2013yxa,Demartin:2014fia,Boudjema:2015nda,Hagiwara:2017ban,Goncalves:2018agy,Faroughy:2019ird,Bortolato:2020zcg,Brod:2023wsh,Miralles:2024huv} attention. As discussed in Refs.~\cite{deVries:2017ncy, DeVries:2018aul, Fuchs:2020uoc,Alonso-Gonzalez:2021jsa,AharonyShapira:2021ize,Bahl:2022yrs}, such effects could be related to one of the most pressing shortcomings of the SM, namely its inability to explain the observed baryon asymmetry of the Universe. Surprisingly, a complex tau Yukawa coupling on its own may already provide a sufficient new source of CP violation beyond the SM in order to successfully account for this, however based on Ref.~\cite{vandeVis:2025efm} this claim needs to be further scrutinized.

The remarkable agreement of the Higgs boson properties with the SM expectations, combined with the absence of any new degrees of freedom at the LHC up to the TeV scale~\cite{CMS:2019ybf, ATLAS:2019fgd}, motivates the use of the SM Effective Field Theory (SMEFT) as a framework to parameterize a large class of short-distance NP effects~\cite{Buchmuller:1985jz,Grzadkowski:2010es,Brivio:2017vri,Isidori:2023pyp}. In this approach, the SM Lagrangian is extended by a tower of higher-dimensional operators $\mc{O}_i$, consistent with the SM gauge symmetry:
\begin{equation}
\label{eq:LSMEFT}
\mc{L}_\sscript{SMEFT}= \sum_i \mc{C}_i \mc{O}_i\,,
\end{equation}
where $\mc{C}_i$ denote the respective Wilson coefficients. The SMEFT has been successfully employed in the context of modified Yukawa couplings of the Higgs boson, see e.g.~\cite{Fuchs:2020uoc,Fajfer:2021cxa,Bahl:2022yrs,Brod:2022bww}.

In this work, we present a comprehensive study of the CP-violating modifications of the leptonic Yukawa couplings, originating both from the SMEFT and from a broad class of simplified extensions of the SM. In Sec.~\ref{sec:EFT_pheno} we perform a phenomenological analysis in the SMEFT, highlighting the complementarity between the LHC measurements and the dipole moments of the leptons, where the Barr-Zee contributions to the electron dipole moment~\cite{Barr:1990vd} play a prominent role. In Sec.~\ref{sec:simplified_extensions}, we take a step toward realistic UV completions by considering a complete set of scalar, fermion, and vector degrees of freedom that can generate CP-violating modifications of the Higgs Yukawa couplings at tree level, in both single- and two-field extensions. Concrete UV realizations typically induce non-trivial correlations in the SMEFT parameter space. Through representative examples, we demonstrate that such realistic UV completions generally lead to phenomenologically important effects, beyond those included in the simplified SMEFT setting. These include lepton dipole moments already at the one-loop level, charged lepton flavor violating effects, and electroweak precision observables at tree level. We conclude in Sec.~\ref{sec:conc}.

\section{EFT phenomenology}
\label{sec:EFT_pheno}
In this section, we parameterize the modifications of the Higgs leptonic Yukawa couplings using the SMEFT at the level of dimension-6 operators. We assume that only a single Wilson coefficient for a specific lepton flavor is modified, thereby affecting exclusively lepton flavor-conserving processes. The bounds are presented on both the real and imaginary parts, derived from direct LHC measurements and complemented by constraints from the leptonic dipole moments.

\subsection{Direct measurements}
\label{subsec:direct}
The effective Lagrangian of the Higgs boson couplings with SM leptons, often used in the context of direct measurements, can be written in the mass basis as\footnote{Throughout this work, the symbol $\ell$ is used with two distinct meanings: in SMEFT operators, it refers to the lepton $\SU(2)_\sscript{L}$ doublet (see Eq.~\eqref{eq:OeH}), while in low-energy expressions, $\ell_i$ denotes the charged lepton field of particular flavor. We trust that the context will prevent any ambiguity.}
\begin{equation}
\label{eq:Lagrkptldeff}
    \mathcal{L}_\sscript{eff} = - \sum_{i} \frac{y_{\ell_i}}{\sqrt{2}}\bar{\ell}_i (\kappa_{\ell_i} + i \tilde{\kappa}_{\ell_i} \gamma_5)\ell_i h \,,
\end{equation}
where $i$ runs over the three lepton flavors $e, \mu, \tau$, with $y_{\ell_i} = \sqrt{2} m_{\ell_i}/v$. The SM limit is recovered by setting $(\kappa_{\ell_i}, \tilde{\kappa}_{\ell_i}) = (1,0)$, and originates from the SM Yukawa Lagrangian term $[Y_\sscript{H}^e]_{ij}\bar{\ell}_i H e_j+\hermc$. While this effective Lagrangian can be UV completed in multiple contexts, in this work we focus on its realization within the SMEFT framework. In particular, the dim-6 operator\footnote{We restrict our analysis to dim-6 SMEFT operators. Higher-dimensional contributions, while possible, are assumed to be subleading and are not included~\cite{Li:2020gnx,Murphy:2020rsh}.}
\begin{equation}
\label{eq:OeH}
   \left[\mathcal{O}_{e H}\right]_{ij} = (\bar\ell_i H e_j) ( H^\dagger  H)\,
\end{equation}
contributes to effective Higgs boson couplings with leptons. After electroweak symmetry breaking~(EWSB) and rotations to the fermion mass basis, we derive the matching relations
\begin{equation}\label{eq:kpkptld}
    \kappa_{\ell_i} = 1- \frac{v^2}{y_{\ell_i}} \re[{\cC}_{e H}]_{ii} \,,
    \quad
    \tilde{\kappa}_{\ell_i} = - \frac{v^2}{y_{\ell_i}} \im[{\cC}_{e H}]_{ii} \,.
\end{equation}
A detailed derivation can be found in e.g. Ref.~\cite{Brod:2022bww}.

\subsubsection{Observables}
The Yukawa couplings in Eq.~\eqref{eq:Lagrkptldeff} modify the CP properties of the Higgs-boson interactions with leptons. These modifications impact the Higgs partial decay rates to leptons via
\begin{equation}
\frac{\Gamma(h\to {\ell_i^+}{\ell_i^-})}{\Gamma(h\to {\ell_i^+}{\ell_i^-})_\sscript{SM}} = \kappa_{\ell_i}^2 + \tilde\kappa_{\ell_i}^2\,,
\end{equation}
and the total width of the Higgs as
\begin{equation}
\frac{\Gamma^h}{\Gamma_\sscript{SM}^h} = 1 + \sum_{i} \left(\frac{\Gamma(h\to {\ell_i^+}{\ell_i^-})}{\Gamma(h\to {\ell_i^+}{\ell_i^-})_\sscript{SM}} - 1\right) \mathrm{BR}(h\to{\ell_i^+}{\ell_i^-})_\sscript{SM} \,,
\end{equation}
where we take the SM branching ratios from Ref.~\cite{Heinemeyer:1559921}. In our phenomenological analysis, we use the measurements of the signal strengths
\begin{equation}\label{eq:signal_strength}
    \mu^{\ell_i} = \frac{\sigma^h}{\sigma^h_{\sscript{SM}}}\frac{ \mathrm{BR}(h\to{\ell_i^+}{\ell_i^-})}{ \mathrm{BR}(h\to{\ell_i^+}{\ell_i^-})_\sscript{SM}}\,,
\end{equation}
as observables. In Eq.~\eqref{eq:signal_strength}, $\sigma^h$ denotes the Higgs production cross section. Note that the experimentally relevant production channels of the Higgs boson are not significantly impacted by the modifications of the leptonic Yukawa couplings, warranting the use of inclusive signal strength measurements. 

In the case of the $\tau$ lepton, further information can be extracted from the correlations of the $\tau$ spin with the momenta of its decay products~\cite{Kramer:1993jn, Berge:2011ij, Berge:2014sra, Berge:2014wta}. A particularly sensitive observable in this context is the angle $\phi_{\sscript{CP}}$ between the $\tau$ decay planes in the Higgs rest frame, which enters the differential decay distribution as~\cite{Berge:2014wta}
\begin{equation}
\label{eq:tauCPVcollider}
    \frac{d\Gamma(h\to \tau^+ \tau^-)}{d\phi_{\sscript{CP}}} \sim 1 - b(E^+) b(E^-) \frac{\pi^2}{16} \cos(\phi_{\sscript{CP}} - 2\alpha^\tau)\,,
\end{equation}
where $b(E^\pm)$ are the spectral functions describing spin analyzing power of a given decay product. Here we have defined the effective mixing angle, which encodes the CP structure of the $h\tau\tau$ interaction as
\begin{equation}
    \tan (\alpha^\tau) = \frac{\tilde{\kappa}_\tau}{\kappa_\tau}\,.
\end{equation}

\subsubsection{Experimental status}

\vspace{0.2cm}
\noindent
$\bm{\tau}$ \textbf{lepton.}  Both ATLAS~\cite{ATLAS:2022akr} and CMS~\cite{CMS:2021sdq} experiments carried out dedicated analyses in order to probe the CP structure of the $\tau$ Yukawa coupling, each based on approximately 140~$\mathrm{fb}^{-1}$ of LHC Run 2 data. These analyses measured the signal strength (combining multiple Higgs production channels) as well as the CP-sensitive angular distribution defined in Eq.~\eqref{eq:tauCPVcollider}. In our analysis, we combine the results of the two collaborations using a multivariate Gaussian approximation of the reported likelihoods. For the signal strength we also take into account ATLAS and CMS measurements of signal strengths on Run 1 data~\cite{ATLAS:2016neq}. We find
\begin{equation}
    \mu^\tau = 1.01 \pm 0.08 \,,
    \qquad
    \alpha^\tau = (5 \pm 13)\,{}^\circ \,,
\end{equation}
with negligible correlation. 

\vspace{0.2cm}
\noindent
\textbf{Muon.} A recent ATLAS analysis combining 165~$\mathrm{fb}^{-1}$ of Run 3 data at 13.6~TeV with Run 2 results has yielded evidence for $h \to \mu^+\mu^-$ decay with a significance of $3.4\sigma$~\cite{ATLAS:2025coj}. The extracted best-fit signal strength is
\begin{equation}
    \mu^\mu_\sscript{ATLAS} = 1.4 \pm 0.4 \,.
\end{equation}
Currently, the available CMS results are limited to Run 2, where evidence was reported with a significance of $3\sigma$~\cite{CMS:2020xwi,CMS:2022dwd}. The measured signal strength is reported as~\cite{CMS:2022dwd}
\begin{equation}
    \mu^\mu_\sscript{CMS} = 1.21^{+0.42}_{-0.38} \mathrm{(stat)}^{+0.17}_{-0.16}\mathrm{(syst)} \,.
\end{equation}
In further analysis we use a combination of the above two measurements:
\begin{equation}
    \mu^\mu_\sscript{comb.} = 1.31 \pm 0.29\,.
\end{equation}

\vspace{0.2cm}
\noindent
\textbf{Electron.} The smallness of the electron Yukawa coupling makes the measurement of $h\to e^+ e^-$ particularly challenging. Nevertheless, dedicated analyses were performed using Run 2 data by both ATLAS~\cite{ATLAS:2019old} and CMS~\cite{CMS:2022urr}, finding the following upper limits on the $h\to e^+ e^-$ branching ratio at the 95\% confidence level (CL) 
\begin{equation}
\label{eq:BrHee}
    \begin{split}
        \mathrm{BR}_{ee}^\sscript{ATLAS} &< 3.6\times 10^{-4} \,,
        \quad
        \mathrm{BR}_{ee}^\sscript{CMS} <  3.0 \times 10^{-4} \,.
    \end{split}
\end{equation}
These bounds are about five orders of magnitude above the SM prediction. We could in principle also consider the constraint on the electron Yukawa from the measurements of the total Higgs decay width. The current PDG world average~\cite{ParticleDataGroup:2024cfk} of the ATLAS~\cite{ATLAS:2023dnm} and CMS~\cite{CMS:2022ley} measurements of the Higgs total width is
\begin{equation}
    \Gamma^h = 3.7^{+1.9}_{-1.4}~\mathrm{MeV} \,.
\end{equation}
However, inferring $\mathrm{BR}_{ee}$ from this quantity leads to a 3 orders of magnitude weaker constraint compared to the bounds reported in Eq.~\eqref{eq:BrHee}.

\subsection{Dipole moments of leptons}
\label{subsec:dipole_moms}

\subsubsection{Observables}
Leptonic electric and magnetic dipole moments (EDMs and MDMs), are described at low energies within the framework of the low-energy effective field theory (LEFT). Below the electroweak scale, their dominant contributions arise from dim-5 dipole operators of the form
\begin{equation}\label{eq:LEFT_dipole}
    \cL_{\sscript{LEFT}}\supset [\cC_{e\gamma}]_{ij}(\bar \ell_{Li} \sig^{\mu\nu}\ell_{{Rj}})F_{\mu\nu}+\hermc\,,
\end{equation}
where $F_{\mu\nu}$ is the electromagnetic field strength tensor, $i,j$ are flavor indices running over the charged leptons $e, \mu, \tau$, and $L,R$ denote the lepton chiralities. At tree level, these LEFT coefficients can be matched onto the SMEFT Wilson coefficients associated with the dim-6 dipole operators:
\begin{equation}\label{eq:SMEFT_dipole_ops}
    \begin{alignedat}{2}
        [\cO_{eB}]_{ij}&=(\bar\ell_{i} \sig^{\mu\nu}e_{j})HB_{\mu\nu}\,,
        \\
        [\cO_{eW}]_{ij}&=(\bar\ell_{i} \sig^{\mu\nu}\tau^Ie_{j})HW^I_{\mu\nu}\,,
    \end{alignedat}
\end{equation}
with the corresponding tree-level matching relation given by~\cite{Jenkins:2017jig}
\begin{equation}
    [\cC_{e\gamma}]_{ij}=\frac{v}{\sqrt2}\Big[ c_{\sscript{W}}[\cC_{eB}]_{ij}-s_{\sscript{W}}[\cC_{eW}]_{ij} \Big]\,.
\end{equation}
Here $v = 246\,\gev$ denotes the vacuum expectation value (VEV) of the Higgs field, while $s_\sscript{W}$ and $c_\sscript{W}$ denote the sine and cosine of the weak mixing angle, respectively. The NP contributions to the EDM and MDM of a given charged lepton $\ell_i$ can be expressed as~\cite{Kley:2021yhn,Guedes:2022cfy}
\begin{equation}
    d_{\ell_i} =-2\,\im[\cC_{e\gamma}]_{ii}\,,
    \quad
    \Delta a_{\ell_i} = -\frac{4m_{\ell_i}}{eQ_\ell} \re[\cC_{e\gamma}]_{ii} \,,
    \vspace{+0.05cm}
\end{equation}
where $Q_\ell=-1$ is the electric charge of the lepton.\footnote{We work in the $D_\mu = \partial_\mu + i e Q_f A_\mu$ convention for the covariant derivative. $a_{\ell_i}$ is independent of this choice, while the sign of the MDM term in the Lagrangian depends on the sign in $D_\mu$.}
In addition, the $\cO_{eH}$ SMEFT operator defined by Eq.~\eqref{eq:OeH} can induce lepton dipole moments at two loops through Barr–Zee type diagrams involving Higgs exchange and heavy fermions or gauge bosons in the loop~\cite{Brod:2022bww,Brod:2013cka,Altmannshofer:2015qra,Davila:2025goc,Abe:2013qla,Cheung:2009fc}. A key feature of this setup is that the electron EDM is generated by a CP-odd interference between the scalar end pseudoscalar terms in the Yukawa couplings. Following the expressions in Ref.~\cite{Brod:2022bww}, the Barr–Zee contributions to the electron EDM can be approximated as
\begin{widetext}
    \begin{equation}\label{eq:BarrZee_semi_numerical}
    \begin{alignedat}{2}
        d_e&=\frac{e^3\,v}{(16\pi^2)^2}
        \Bigg\{\lzs (3.68\pm 0.92)\log\frac{m_H^2}{\Lambda^2}-7.68 \dzs\im[\cC_{eH}]_{11}
        -1.99\times10^{\eminus6}\,\im[\cC_{eH}]_{22}
        -1.24\times10^{\eminus5}\,\im[\cC_{eH}]_{33}
        \Bigg\}\,,
    \end{alignedat}
\end{equation}
\end{widetext}
where $\Lambda$ denotes the UV scale. Several comments are in order regarding Eq.~\eqref{eq:BarrZee_semi_numerical}. First, the electron EDM receives contributions from Higgs couplings to muons and taus, which stem from the fact that heavier charged leptons can circulate in the internal loop of the Barr–Zee diagrams, thereby inducing terms proportional to $\im[\cC_{eH}]_{22}$ and $\im[\cC_{eH}]_{33}$, in addition to the dominant electron term. Second, the numerical prefactors multiplying each contribution originate from the two-loop Barr–Zee loop functions, which depend on the mass of the internal fermion as well as the topology of the Higgs and gauge boson interactions. Finally, the $\log$ term captures the leading-$\log$ effect due to the two-loop mixing in the SMEFT, while the rest of the terms are due to finite matching contributions.

The contributions of the modified electron Yukawa coupling to the anomalous magnetic moment of the electron can be obtained via the relation~\cite{Altmannshofer:2015qra}
\begin{equation}
    \Delta a_{e} = -\frac{2m_e}{e}\,\sum_i \frac{\re[C_{eH}]_{ii}}{\im[C_{eH}]_{ii}}\,d_{e}\big|_{\im[C_{eH}]_{ii}}\,, 
\end{equation}
where $d_{e}|_{\im[C_{eH}]_{ii}}$ is given by Eq.~\eqref{eq:BarrZee_semi_numerical} with $\im[C_{eH}]_{ii}$ as the only non-vanishing Wilson coefficient. For the Barr-Zee contributions to the electric and magnetic dipole moments of the muon we employ similar formulas based on the results presented in Refs.~\cite{Brod:2022bww, Altmannshofer:2015qra}.

Finally, we briefly comment on the impact of one-loop renormalization group equation (RGE) effects. If one starts with the $\cO_{eH}$ operator (see Eq.~\eqref{eq:OeH}) defined at a high scale in the SMEFT, gauge interactions induce an $\sim 5\%$ correction to the corresponding coefficient when running from $\mu_i\sim 1\,\tev$ down to the electroweak scale $\mu_f\sim100\,\gev$, within the leading-log approximation. After matching onto the LEFT, the corresponding dipole operator undergoes an additional $\sim 4\%$ modification when evolved further to $\mu_f\sim2\,\gev$~\cite{Alonso:2013hga,Jenkins:2017dyc}. Since these RGE effects remain at the few-percent level compared to the overall size of the constraints, we consistently omit them in the discussion below.

\subsubsection{Experimental status}
\label{sec:dipole_exp_status}
\noindent
\textbf{EDMs.} The electron EDM is the most stringently constrained, thanks to the highly precise measurements performed in polar molecules, which amplify the effects of an EDM through strong internal electric fields. The most recent measurement, performed using HfF$^{+}$ molecular ions, sets a limit of~\cite{Roussy:2022cmp}
\begin{equation}
    |d_e|<4.1\times 10^{\eminus30}\, e \,\mathrm{cm}
    \qquad90\%\,\mathrm{CL}\,,
\end{equation}
where we neglect the electron-nucleon couplings~\cite{Chupp:2017rkp, Ardu:2025rqy}.

The situation is more challenging for the heavier leptons, where experimental sensitivity remains significantly weaker. For the muon, EDM searches are typically conducted in storage ring experiments, where spin precession in combined electric and magnetic fields can reveal EDM-induced effects. The best experimental upper limit is~\cite{Muong-2:2008ebm} 
\begin{equation}
    |d_\mu|<1.9\times 10^{\eminus19}\,e\, \mathrm{cm}
    \qquad 95\%\,\mathrm{CL}\,,
\end{equation}
which is several orders of magnitude weaker than bound for the electron. A slightly improved limit can be obtained from atomic and molecular EDMs~\cite{Ema:2021jds}.

Lastly, in case of the tau lepton, only indirect bounds exist due to its short lifetime, which prevents the use of conventional techniques such as spin precession for direct EDM measurements. The strongest existing limits on $d_\tau$ arise from precision measurements of tau polarizations in the process $e^+ e^- \to \tau^+ \tau^-$ at Belle. 
The current best bound reads~\cite{Belle:2021ybo,Escribano:1996wp,Belle:2002nla} 
\begin{equation}
    -1.85< \frac{d_\tau}{10^{-17} e\,\mathrm{cm}} < 0.61
    \qquad 95\%\,\mathrm{CL}\,,
\end{equation}
and will not be used in the phenomenological analyses in this work.

\noindent
\textbf{MDMs.} The magnetic moments of the electron and muon have been both predicted and measured to a very high precision, and present a cornerstone test of the quantum structure of the SM. The electron MDM within the SM is dominated by the quantum electrodynamics contributions. Perturbative calculations of $a_e$ which include terms up to $\mc{O}(\alpha^5)$ have been performed~\cite{Aoyama:2017uqe,Aoyama:2019ryr}. 
From the measurements of $ a_e$~\cite{ParticleDataGroup:2024cfk,Fan:2022eto,Hanneke:2008tm}
\begin{equation}
   a_e^\sscript{exp} = (1159.65218062 \pm 0.00000012)\E{-6}\,,     
\end{equation}
we typically infer the most precise value of the fine structure constant $\alpha$. However, in our case we use $a_e^\sscript{exp}$ as a constraint, and so we have to utilize an independent measurement of $\alpha$ in order to predict $a_e^\sscript{SM}$. The second most precise determination follows from atom interferometry. For the Cs atom the extracted value of $\alpha^{-1}_\sscript{Cs}$ and the corresponding prediction of $a_e^\sscript{SM}$ are~\cite{Parker:2018vye,Aoyama:2019ryr}:
\begin{equation}
\label{eq:alphaCs}
\begin{split}
    \alpha^{-1}_\sscript{Cs} &= 137.035999046(27)\,,\\
    a_e^\sscript{SM,Cs} &= 1159.652181606(230)\E{-6}\,.
    \end{split}
\end{equation}
The difference between the experimental value and the SM prediction of the electron MDM is
\begin{equation}
    \label{eq:deltaae}
   \Delta a_e^\sscript{Cs} = a_e^\sscript{exp} - a_e^\sscript{SM,Cs} = 
   (-0.99\pm 0.26)\E{-12}\,,
\end{equation}
showcasing a tension at the level of $3.8\sigma$. However, the experimental status of $\alpha$ in atom interferometry measurements is unclear at the moment, since the measurement on rubidium atoms~\cite{Morel:2020dww,ParticleDataGroup:2024cfk} results in SM prediction, $a_e^\sscript{SM,Rb} = 1159.652180252(95)\E{-6}$, which is $5.5\sigma$ lower than the $a_e^\sscript{SM,Cs}$ of Eq.~\eqref{eq:alphaCs}. There is no straightforward way to average $a_e^\sscript{SM,Cs}$ and $a_e^\sscript{SM,Rb}$ and we choose to employ the result based on caesium, Eq.~\eqref{eq:deltaae}, as our experimental constraint. Note that using $\alpha_\sscript{Rb}$ the discrepancy lowers to $2.4\sigma$ and changes sign: $\Delta a_e^\sscript{Rb} = (0.37 \pm 0.15)\E{-12}$.

The SM prediction for the muon anomalous magnetic moment has shifted significantly once state-of-the art lattice QCD studies of the hadronic vacuum polarization have been taken into account~\cite{Borsanyi:2020mff,Djukanovic:2024cmq,RBC:2024fic,Boccaletti:2024guq,Aliberti:2025beg, Athron:2025ets}:
\begin{equation}
 a_\mu^\sscript{SM} = 116592033(62)\E{-11}\,.
\end{equation}
The SM prediction can be compared to the latest experimental value~\cite{Aliberti:2025beg,Muong-2:2025xyk}
\begin{equation}
 a_\mu^\sscript{exp} = (116592 071.5\pm14.5)\E{-11}\,,
\end{equation}
in order to find that the discrepancy between the two is consistent with zero~\cite{Aliberti:2025beg}:
\begin{equation}
    \Delta a_\mu = a_\mu^\sscript{exp} - a_\mu^\sscript{SM} = 38(63)\E{-11}\,.
\end{equation}

Finally, we comment on the anomalous magnetic moment of the $\tau$ lepton. Similarly to its EDM, the measurement of $a_\tau$ proves to be experimentally very challenging. The current best constraint comes from Pb+Pb collisions at the LHC via the $\gamma\gamma \to \tau^+\tau^-$ process, where ATLAS measured the 95\% CL interval of~\cite{ATLAS:2022ryk}
\begin{equation}
    -0.057 < a_\tau < 0.024\,.
\end{equation}
While this result is compatible with the SM expectation~\cite{Eidelman:2007sb}, it will not be used in the subsequent phenomenological analyses.

\begin{figure*}[t]
  \centering\begin{tabular}[t]{p{0.33\linewidth}p{0.33\linewidth}p{0.33\linewidth}}
        \includegraphics[scale=0.56]{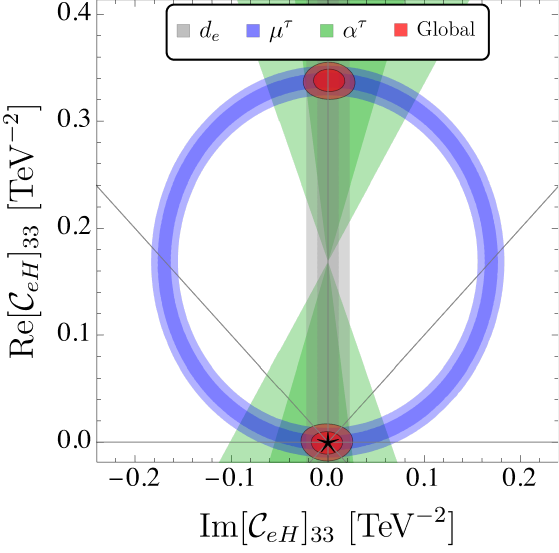} & 
        \includegraphics[scale=0.565]{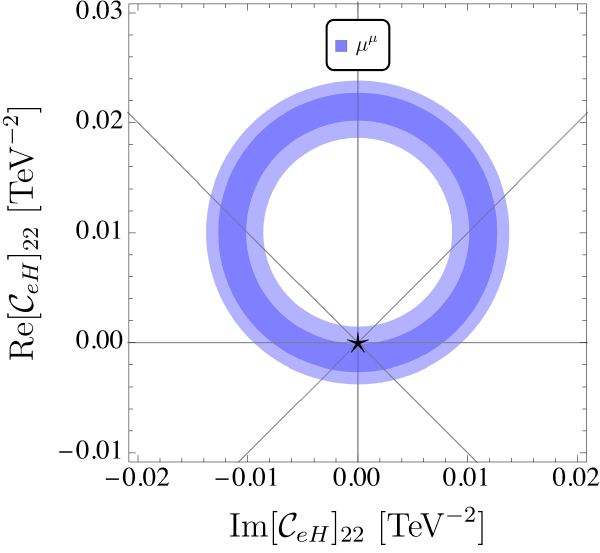} &
         \includegraphics[scale=0.535]{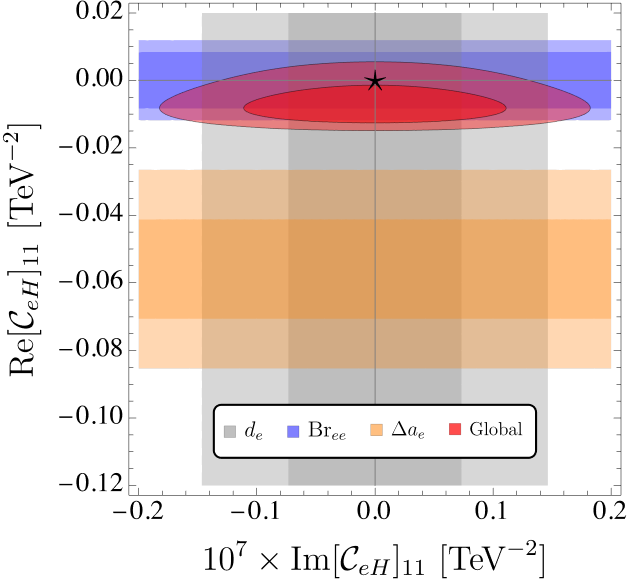} 
    \end{tabular}
    \caption{The $68\%$ CL (dark) and $95\%$ CL (light) constraints on the real and imaginary part of the $\cC_{eH}$ Wilson coefficient under three flavor assumptions: $\tau$-specific (left plot), $\mu$-specific (middle plot) and electron-specific (right plot). See Sec.~\ref{sec:EFT_results} for a discussion.}
    \label{fig:CeH}
\end{figure*}

\subsection{EFT results}
\label{sec:EFT_results}
Here we analyze the constraining power of the above-mentioned observables on the diagonal components of $\cC_{eH}$, assuming a presence of a single lepton flavored Wilson coefficient at a time.

\vspace{0.2cm}
\noindent
$\bm{\tau}$ \textbf{lepton.} We show the constraints on $[\cC_{eH}]_{33}$ in Fig.~\ref{fig:CeH} (left), plotting the constraints from $pp\to h \to \tau^+\tau^-$ signal strength measurements and angular analyses discussed in Sec.~\ref{subsec:direct}, and from the Barr-Zee contribution of $\im[\cC_{eH}]_{33}$ to the electron EDM discussed in Sec.~\ref{subsec:dipole_moms}. As expected the signal strength measurement alone is insensitive to the phase of $\cC_{eH}$, while the measurements of the angular distributions from Eq.~\eqref{eq:tauCPVcollider} break the degeneracy. Note that the $\alpha^\tau$ contours are slightly asymmetric in the imaginary direction, which is possible due to the CP-sensitive nature of the observable. However, with the current precision, the Barr-Zee effect is stronger and leads to the dominant constraint in the $\im[\cC_{eH}]_{33}$ direction. In red, we show the two degenerate minima of the global fit: the lower one is consistent with the SM expectations, while the upper one fine-tunes the dim-4 and dim-6 contributions to the tau Yukawa.

\vspace{0.2cm}
\noindent
\textbf{Muon.} We show the constraints on $[\cC_{eH}]_{22}$ in Fig.~\ref{fig:CeH} (middle). Note that we only show the constraints from $pp\to h \to \mu^+\mu^-$ signal strength measurements discussed in Sec.~\ref{subsec:direct}. Contrary to the $\tau$ case, there are no dedicated collider angular analyses in the case of muons. The Barr-Zee contributions to the electron EDM and the muon MDM both lead to less sensitive constraints in the plane of $[\cC_{eH}]_{22}$ by about an order of magnitude.

\vspace{0.2cm}
\noindent
\textbf{Electron.} Finally, we show the constraints in the complex plane of $[\cC_{eH}]_{11}$ on Fig.~\ref{fig:CeH} (right). The upper limits on the branching ratio of $h\to e^+e^-$ by CMS and ATLAS yield currently the best constraints in the real direction. For comparison purposes, we show the constraints due to Barr-Zee contributions to the electron MDM in orange, showcasing the tension with the SM prediction. Nevertheless, this constraint is currently not particularly competitive in this plane. Orthogonally, the constraint due to Barr-Zee induced electron EDM leads to a remarkably tight constraint on the imaginary component of $[\cC_{eH}]_{11}$.

\begin{table*}[t]
\centering
\scalebox{0.875}{
\begin{tabular}{c@{\hspace{1.5cm}}c@{\hspace{0.7cm}}c@{\hspace{0.7cm}}c}
\toprule
\textbf{Type}
&\textbf{Field 1} 
&\textbf{Field 2}
&$\bm{-\cL_{\sscript{UV}}\supset}$ 
\\
\midrule
$\bm{({S}_1,{S}_2)}$
&$\cS\sim(\rep1,\rep1)_0$
&$\varphi\sim(\rep1,\rep2)_{1/2}$
&$\kappa_\cS\,\cS H^\dag H+\Big[ [Y^e_\varphi]_{ij}\varphi\,\bar\ell_ie_j+\kappa_{\cS\varphi}\cS\varphi^\dag H+\hermc \Big]
		$
\\[3.4pt]
$\bm{({S}_1,{S}_2)}$
&$\Xi\sim(\rep1,\rep3)_0$
&$\varphi\sim(\rep1,\rep2)_{1/2}$
&$\kappa_\Xi\, H^\dag \Xi^a\sig^a H
		+\Big[ [Y^e_\varphi]_{ij}\varphi\,\bar\ell_ie_j+\kappa_{\Xi\varphi}\Xi^a(\varphi^\dag \sig^a H)+\hermc \Big]$
\\[3.4pt]
\multirow{1}{*}{\vspace{-0.0cm}$\bm{({S}_1,{S}_2)}$}
&\multirow{1}{*}{\vspace{-0.0cm}$\Xi_1\sim(\rep1,\rep3)_1$}
&\multirow{1}{*}{\vspace{-0.0cm}$\varphi\sim(\rep1,\rep2)_{1/2}$}
&$\kappa_{\Xi_1} \Xi_{1}^{a\dag}(\tilde\phi^\dag\sig^a\phi)+[y_{\Xi_1}]_{ij}\Xi_{1}^{a\dag}\bar\ell_{i}\sig^ai\sig_2\ell_{j}^c+ [Y^e_\varphi]_{ij}\varphi\,\bar\ell_ie_j
+\kappa_{\Xi_1\varphi}\Xi_1^{a\,\dag}(\tilde\varphi^\dag \sig^a H)+\hermc$
\\[2.6pt]
\midrule
$\bm{(F_1,F_2)}$
&$E\sim(\rep1,\rep1)_{\eminus1}$
&$\Delta_1\sim(\rep1,\rep2)_{\eminus1/2}$
&$[\lambda_E]_{i}\bar E_{R}H^\dag\ell_{i}+[\lambda_{\Delta_1}]_{i}\bar\Delta_{1L}H e_{i}+\lambda_{E\Delta_1}\bar E_L H^\dag \Delta_{1R}+\hermc$
\\[2.6pt]
$\bm{(F_1,F_2)}$
&$E\sim(\rep1,\rep1)_{\eminus1}$
&$\Delta_3\sim(\rep1,\rep2)_{\eminus3/2}$
&$[\lambda_E]_{i}\bar E_{R}H^\dag\ell_{i}+[\lambda_{\Delta_3}]_{i}\bar\Delta_{3L}\tilde H e_{i}+\lambda_{E\Delta_3}\bar E_L\tilde H^\dag\Delta_{3R}+\hermc$
\\[2.6pt]
$\bm{(F_1,F_2)}$
&$\Sigma\sim(\rep1,\rep3)_0$
&$\Delta_1\sim(\rep 1,\rep2)_{\eminus1/2}$
&$\frac{1}{2}[\lambda_\Sigma]_{i}\bar\Sigma^a_{R}\tilde H^\dag\sig^a\ell_{i}+[\lambda_{\Delta_1}]_{i}\bar\Delta_{1L}H e_{i}+\frac{1}{2}\lambda_{\Sigma\Delta_1}\bar\Sigma^{c\,a}_R\tilde H^\dag \sigma^a \Delta_{1R}+\hermc$
\\[2.6pt]
$\bm{(F_1,F_2)}$
&$\Sigma_1\sim(\rep1,\rep3)_{\eminus1}$
&$\Delta_1\sim(\rep 1,\rep2)_{\eminus1/2}$
&$\frac{1}{2}[\lambda_{\Sigma_1}]_{i}\bar\Sigma^a_{1R}H^\dag\sig^a\ell_{i}+[\lambda_{\Delta_1}]_{i}\bar\Delta_{1L}H e_{i}+\frac{1}{2}\bar\Sigma^a_{1L}H^\dag\sigma^a\Delta_{1R}+\hermc$
\\[2.6pt]
$\bm{(F_1,F_2)}$
&$\Sigma_1\sim(\rep1,\rep3)_{\eminus1}$
&$\Delta_3\sim(\rep 1,\rep2)_{\eminus3/2}$
&$\frac{1}{2}[\lambda_{\Sigma_1}]_{i}\bar\Sigma^a_{1R}H^\dag\sig^a\ell_{i}+[\lambda_{\Delta_3}]_{i}\bar\Delta_{3L}\tilde H e_{i}+\frac{1}{2}\lambda_{\Sigma_1\Delta_3}\bar\Sigma^a_{1L}\tilde H^\dag \sigma^a \Delta_{3R}+\hermc$
\\[2.6pt]
\midrule
$\bm{(S,F)}$
&$\cS\sim(\rep1,\rep1)_0$
&$E\sim(\rep1,\rep1)_{\eminus1}$
&$\kappa_\cS\,\cS H^\dag H+\Big[[\lambda_E]_{i}\bar E_{R}H^\dag\ell_{i}+[\lambda_{\cS E}]_i \cS \bar E_L e_i+\hermc\Big]$
\\[2.6pt]
$\bm{(S,F)}$
&$\cS\sim(\rep1,\rep1)_0$
&$\Delta_1\sim(\rep 1,\rep2)_{\eminus1/2}$
&$\kappa_\cS\,\cS H^\dag H+\Big[[\lambda_{\Delta_1}]_{i}\bar\Delta_{1L}H e_{i}+[\lambda_{\cS\Delta_1}]_i \cS\bar\Delta_{1L}\ell_i+\hermc\Big]$
\\[2.6pt]
$\bm{(S,F)}$
&$\Xi\sim(\rep1,\rep3)_0$
&$\Delta_1\sim(\rep 1,\rep2)_{\eminus1/2}$
&$\kappa_\Xi\, H^\dag \Xi^a\sig^a H+\Big[[\lambda_{\Delta_1}]_{i}\bar\Delta_{1L}H e_{i}+[\lambda_{\Xi\Delta_1}]_i \Xi^a \bar\Delta_{1R}\sig^a\ell_i+\hermc\Big]$
\\[2.6pt]
$\bm{(S,F)}$
&$\Xi\sim(\rep1,\rep3)_0$
&$\Sigma_1\sim(\rep1,\rep3)_{\eminus1}$
&$\kappa_\Xi\, H^\dag \Xi^a\sig^a H
		+\Big[\frac{1}{2}[\lambda_{\Sigma_1}]_{i}\bar\Sigma^a_{1R}H^\dag\sig^a\ell_{i}+[\lambda_{\Xi\Sigma_1}]_i\Xi^a \bar\Sigma^a_{1L}e_i+\hermc\Big]$
\\[2.6pt]
\multirow{1}{*}{\vspace{-0.0cm}$\bm{(S,F)}$}
&\multirow{1}{*}{\vspace{-0.0cm}$\Xi_1\sim(\rep1,\rep3)_1$}
&\multirow{1}{*}{\vspace{-0.0cm}$\Delta_3\sim(\rep 1,\rep2)_{\eminus3/2}$}
&$\kappa_{\Xi_1} \Xi_{1}^{a\dag}(\tilde\phi^\dag\sig^a\phi)+[y_{\Xi_1}]_{ij}\Xi_{1}^{a\dag}\bar\ell_{i}\sig^ai\sig_2\ell_{j}^c+[\lambda_{\Delta_3}]_{i}\bar\Delta_{3L}\tilde H e_{i}+[\lambda_{\Xi_1\Delta_3}]_i \Xi^{a\,\dag}_1\bar\Delta_{3R}\sig^a\ell_i+\hermc$
\\[2.6pt]
\multirow{1}{*}{\vspace{-0.0cm}$\bm{(S,F)}$}
&\multirow{1}{*}{\vspace{-0.0cm}$\Xi_1\sim(\rep1,\rep3)_1$}
&\multirow{1}{*}{\vspace{-0.0cm}$\Sigma\sim(\rep1,\rep3)_0$}
&$\kappa_{\Xi_1} \Xi_{1}^{a\dag}(\tilde\phi^\dag\sig^a\phi)+[y_{\Xi_1}]_{ij}\Xi_{1}^{a\dag}\bar\ell_{i}\sig^ai\sig_2\ell_{j}^c+\frac{1}{2}[\lambda_\Sigma]_{i}\bar\Sigma^a_{R}\tilde H^\dag\sig^a\ell_{i}+[\lambda_{\Xi_1\Sigma}]_i \Xi_1^{a\,\dag}\bar\Sigma^{c\,a}_Re^c_i+\hermc$
\\[2.6pt]
\bottomrule
\end{tabular}
}	
\caption{Overview of the two-field UV configurations analyzed in this work. The first column indicates the type of two-field setup. The second and third columns list the field content along with the corresponding SM gauge quantum representations. The final column collects the relevant UV interaction terms.}
\label{tab:config_summary}
\end{table*}

\begin{table*}[t]
\centering
\scalebox{1.00}{
\begin{tabular}{c@{\hspace{3.4cm}}c@{\hspace{3.4cm}}c@{\hspace{3.4cm}}c}
\toprule
$\bm{(\mathcal F_1,\mathcal F_2)}$ 
&$\bm{\alpha(\mathcal F_1,\mathcal F_2)}$
&$\bm{\beta(\mathcal F_1,\mathcal F_2)}$
&$\bm{X(\mathcal F_1,\mathcal F_2)}$
\\
\midrule
$(\cS,\varphi)$
&$-1$
&-
&$\kappa_{\cS\varphi}\kappa_\cS$
\\[2.8pt]
$(\Xi,\varphi)$
&$-1$
&-
&$\kappa_{\Xi\varphi}\kappa_\Xi$
\\[2.8pt]
\multirow{1}{*}{$(\Xi_1,\varphi)$}
&$-2$
&-
&$\kappa_{\Xi_1\varphi}^*\kappa_{\Xi_1}$
\\[2pt]
\midrule
$(E,\Delta_1)$
&$-1$
&$-1/4$
&$\lambda_{E\Delta_1}[\lambda_E]^*_i[\lambda_{\Delta_1}]_j$
\\[2.5pt]
$(E,\Delta_3)$
&$-1$
&$-5/4$
&$\lambda_{E\Delta_3}[\lambda_E]^*_i[\lambda_{\Delta_3}]_j$
\\[2.5pt]
$(\Sigma,\Delta_1)$
&$-1/2$
&$-1/8$
&$\lambda_{\Sigma\Delta_1}[\lambda_\Sigma]^*_i[\lambda_{\Delta_1}]_j$
\\[2.5pt]
$(\Sigma_1,\Delta_1)$
&$-1/4$
&$-9/16$
&$\lambda_{\Sigma_1\Delta_1}[\lambda_{\Sigma_1}]^*_i[\lambda_{\Delta_1}]_j$
\\[2.5pt]
$(\Sigma_1,\Delta_3)$
&$+1/4$
&$+5/16$
&$\lambda_{\Sigma_1\Delta_3}[\lambda_{\Sigma_1}]^*_i[\lambda_{\Delta_3}]_j$
\\[2.5pt]
\midrule
$(\cS,E)$
&$-1$
&$-1/12$
&$\kappa_\cS [\lambda_E]^*_i[\lambda_{\cS E}]_j$
\\[2.5pt]
$(\cS,\Delta_1)$
&$-1$
&$-1/12$
&$\kappa_\cS [\lambda_{\cS\Delta_1}]^*_i[\lambda_{\Delta_1}]_j$
\\[2.5pt]
$(\Xi,\Delta_1)$
&$-1$
&$+1/4$
&$\kappa_\Xi [\lambda_{\Xi\Delta_1}]^*_i[\lambda_{\Delta_1}]_j$
\\[2.5pt]
$(\Xi,\Sigma_1)$
&$-1/2$
&$-3/8$
&$\kappa_\Xi[\lambda_{\Sigma_1}]^*_i[\lambda_{\Xi\Sigma_1}]_j$
\\[2.5pt]
\multirow{1}{*}{$(\Xi_1,\Delta_3)$}
&$-2$
&$-1$
&$\kappa_{\Xi_1}[\lambda_{\Xi_1\Delta_3}]^*_i[\lambda_{\Delta_3}]_j$
\\[2.5pt]
\multirow{1}{*}{$(\Xi_1,\Sigma)$}
&$-1$
&$\,0$
&$\kappa_{\Xi_1}[\lambda_{\Sigma_1}]^*_i[\lambda_{\Xi_1\Sigma}]_j^*$
\\[2pt]
\bottomrule
\end{tabular}

}	
\caption{Summary of numerical prefactors and UV coupling structures for all two-field configurations considered in the analysis. The first column lists the pairs of mediator fields $(\mathcal{F}_1, \mathcal{F}_2)$, while the second and third columns give the coefficients $\alpha(\mathcal{F}_1, \mathcal{F}_2)$ and $\beta(\mathcal{F}_1, \mathcal{F}_2)$ that enter the expressions for the tree-level Wilson coefficient $\cC_{eH}$ (see Eq.~\eqref{eq:CeH_two_field_config}) and the one-loop dipole coefficient $\cC_{e\gamma}$ (see Eq.~\eqref{eq:dipole_2_field_gen}), respectively. The last column provides the flavor structure $X_{ij}(\mathcal{F}_1, \mathcal{F}_2)$, built from the relevant UV couplings for each configuration. See Tab.~\ref{tab:config_summary} for the overview of the two-field configurations.}
\label{tab:2field_matching_params}
\end{table*}
\section{Simplified extensions of the SM}
\label{sec:simplified_extensions}

As demonstrated in the EFT framework developed in Sec.~\ref{sec:EFT_pheno}, CP-violating modifications of the Higgs Yukawa couplings are particularly compelling due to the complementarity between collider and low-energy constraints. Here we discuss classes of simplified extensions of the SM, consisting of one or two new heavy fields, each taken to be a scalar, fermion, or vector, and transforming under different representations of the SM gauge group. Such extensions have been systematically classified in Ref.~\cite{deBlas:2017xtg} in context of the complete set of dim-6 SMEFT operators obtained after integrating them out at tree level. Our baseline criterion for models of interest is nonetheless deliberately stringent: they should be capable of generating complex $\cC_{eH}$ at tree level, thereby maximizing the discovery potential in collider probes of CPV Higgs Yukawa couplings. In what follows, we focus exclusively on the interaction terms that directly induce complex diagonal components of $\cC_{eH}$ at tree level, setting all other couplings to zero.

In the following subsections, we systematically examine such single- and two-field configurations, paying particular attention to the phenomenologically relevant complementary operators they generate. Among these are the lepton dipole operators, which arise either directly at one loop or via two-loop Barr-Zee-type contributions. Moreover, depending on the flavor structure of the UV couplings, many scenarios also give rise to the operators contributing to charged lepton flavor-violating (cLFV) transitions such as $\ell_i\to \ell_j\gamma$ or $\ell_i\to 3\ell_j$. Lastly, in addition to these observables, we also account for electroweak precision tests (EWPT) and differential cross section measurements at LEP-II. To illustrate the phenomenological importance of the constraints discussed above, we conclude this section with a set of representative scenarios defined by a variety of specific flavor assumptions, ranging from scenarios coupling to single lepton flavors at a time, to two-lepton, universal and minimal flavor violation (MFV) scenarios.

\subsection{Matching}
\label{sec:Matching}
We begin by presenting the SMEFT matching relations for both single- and two-field extensions of the SM considered in this work. These form the basis for the phenomenological analysis presented in the subsequent sections.

\subsubsection{Single-field extensions}
\label{sec:single_field_extensions}
Despite the ability of many single-field extensions to generate the modified Yukawa operators, the configurations of this type are subject to strong structural constraints. As a representative example, let us consider a fermion field $\Delta_1\sim(\rep1,\rep2)_{-1/2}$, whose interactions are listed in Tab.~\ref{tab:config_summary}. The tree-level matching for $\cC_{eH}$ takes the form~\cite{deBlas:2017xtg,Fuentes-Martin:2022jrf}
\begin{equation}
\label{eq:CeH_Delta1}
    [\cC_{eH}]_{ij}=\frac{[Y^e_{\sscript{H}}]_{ik}}{2M^2_{\Delta_1}}[\lambda_{\Delta_1}]_j[\lambda_{\Delta_1}]^*_k\,,
\end{equation}
where $Y_\sscript{H}^e$ denotes the Higgs Yukawa matrix of the leptons, defined under Eq.~\eqref{eq:Lagrkptldeff}. Evidently, the Wilson coefficient depends on the product of a coupling and its complex conjugate. As a result, the diagonal components $i=j$, which we are particularly interested in, are real and can not represent new sources of CP violation in the modified Higgs Yukawa couplings.

On the other hand, for extensions involving vector bosons, the modification to the Yukawa sector stems from interactions between the vector and the Higgs field of the form $g_\mathcal{V}^H\mathcal V^\mu H^\dag iD_\mu H$. In UV-complete models where the vector boson originates from a spontaneously broken gauge symmetry, such interactions arise from the covariant derivative and are thus fixed by the gauge coupling. As a result, the $g_\mathcal{V}^H$ coupling is necessarily real, implying that this class of extensions cannot induce CP-violating contributions to the Yukawa sector. 

The only single-field extension satisfying our baseline criterion is the scalar $\varphi\sim(\rep1,\rep2)_{-1/2}$, transforming under the same gauge representation as the SM Higgs doublet.\footnote{Scalar extensions such as $\varphi$ can arise in frameworks like Two-Higgs-Doublet Models (2HDMs)~\cite{Anisha:2023vvu,Davila:2025goc,Abe:2013qla,Ilisie:2015tra}, where both doublets typically participate in electroweak symmetry breaking. It is worth noting, however, that for generic 2HDMs the lowest-dimension SMEFT or Higgs EFT expansion does not capture the full phenomenology~\cite{Dawson:2023ebe,Banta:2023prj}. In our analysis we consider instead a simplified setup in which $\varphi$ does not acquire a vacuum expectation value and is treated as a heavy state integrated out above the electroweak scale, giving rise to effective operators within the SMEFT framework.} In this setup, renormalizable quartic interactions involving $\varphi$ and three SM Higgs fields can be constructed, along with new Yukawa-like couplings between $\varphi$ and the SM leptons:\footnote{While the $\varphi$ scalar may also couple to SM quarks through Yukawa-like interactions, we set these to zero throughout our analysis. We likewise neglect portal terms such as $\varphi^2 H^2$ and scalar self-couplings of the form $\varphi^4$, as they do not impact the observables under consideration.}
\begin{equation}
\label{eq:varphiint}
    -\cL_{\sscript{UV}}\supset\lambda_\varphi(\varphi^\dag H)(H^\dag H)
    +[Y_\varphi^e]_{ij}\bar\ell_i\, \varphi\, e_j
    +\hermc\,.
\end{equation}
These ingredients allow for a tree-level modification of the Higgs-lepton Yukawa coupling, yielding a Wilson coefficient of the form 
\begin{equation}
\label{eq:varphiCeH}
[\cC_{eH}]_{ij}=\frac{\lambda_\varphi [Y_\varphi^e]_{ij}}{M_\varphi^2} \,,
\end{equation}
which can carry a complex phase and thus induce CP violation even in flavor-conserving transitions ($i=j$).

On the other hand, the dipole operators are generated at one-loop level by $\varphi$. We derive the corresponding Wilson coefficient $[\cC_{e\gamma}]_{ij}$ using \texttt{Matchete}~\cite{Fuentes-Martin:2022jrf} as
\begin{equation}\label{eq:dipole_op_varphi}
    \begin{alignedat}{2}
        [\cC_{e\gamma}]_{ij}=-\frac{e}{384\pi^2}\frac{v}{\sqrt2}&\frac{1}{M_\varphi^2}
        \Big[
        2[Y_\varphi^e Y_\varphi^{e\dagger}Y_\sscript{H}^e]_{ij}+[Y_\sscript{H}^e Y_\varphi^{e\dagger}Y_\varphi^e]_{ij}
        \\&
        ~~~~-12[Y_\varphi^e]_{ij}\mathrm{Tr}[Y_\varphi^{e\dagger}Y_\sscript{H}^e]
        \Big]\,,
    \end{alignedat}
\end{equation}
where we again used the Higgs Yukawa matrix $Y_\sscript{H}^e$ defined under Eq.~\eqref{eq:Lagrkptldeff}. Note that only the last term can lead to potential CP-violating effects for the flavor diagonal ($i=j$) case.

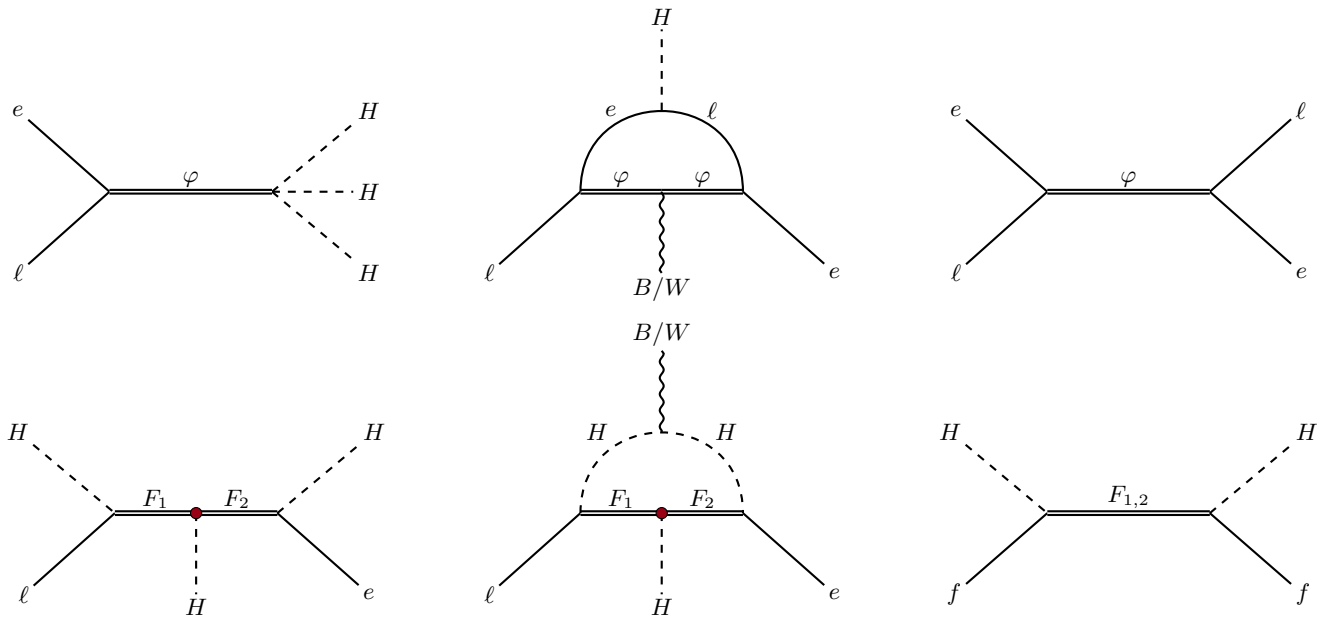
\begin{figure*}[t]
\centering
\centering
\begin{tabular}{c@{\hspace{1.2cm}}c@{\hspace{1.2cm}}c}
\begin{tikzpicture}[baseline=(a2.base), scale=0.54, transform shape]
			\begin{feynman}
			\vertex (a2);
			\vertex [above =2.0cm of a2](a2A);
			\vertex [below =2.0cm of a2](a2B);
			\vertex [left =2.0cm of a2A](a2AL) {\(\scalebox{1.85}{$e$}\)};
			\vertex [left =2.0cm of a2B](a2BL) {\(\scalebox{1.85}{$\ell$}\)};
			\vertex [right =2.0cm of a2](a2M) ;
			\vertex [right =2.0cm of a2M](a2MM);
			\vertex [above =2.0cm of a2MM](a2MMA);
			\vertex [below =2.0cm of a2MM](a2MMB);
			\vertex [right =2.0cm of a2MMA](a2MMAL) {\(\scalebox{1.85}{$H$}\)};
			\vertex [right =2.0cm of a2MMB](a2MMBL) {\(\scalebox{1.85}{$H$}\)};
			\vertex [below =2.0cm of a2M](a2MDB) ;
			\vertex [right =2.0cm of a2MM](a2MMright) {\(\scalebox{1.85}{$H$}\)};
			\diagram*{
			(a2AL)--[plain, thick](a2)--[plain, thick](a2BL),
			(a2MM)--[dashed, thick](a2MMAL),
			(a2MM)--[dashed, thick](a2MMBL),
			(a2MM)--[dashed, thick](a2MMright),
			(a2)--[double, thick, edge label=\(\scalebox{1.85}{$\varphi$}\)](a2MM)
			};
			\draw[fill=red] (a2M) circle(0mm);
			\end{feynman}
	\end{tikzpicture}
&
\begin{tikzpicture}[baseline=(a2.base), scale=0.54, transform shape]
			\begin{feynman}
			\vertex (a1);
			\vertex [below =2.0cm of a2](a2B) ;
			\vertex [left =2.0cm of a2B](a2BL) {\(\scalebox{1.85}{$\ell$}\)};
			\vertex [right =2.0cm of a2](a2M) ;
			\vertex [right =2.0cm of a2M](a2MM);
			\vertex [below =2.0cm of a2MM](a2MMB) ;
			\vertex [right =2.0cm of a2MMB](a2MMBR) {\(\scalebox{1.85}{$e$}\)};
			\vertex [below =2.0cm of a2M](a2MDB) {\(\scalebox{1.85}{$B/W$}\)};
			\vertex [above =2.0cm of a2M](a2MDA) ;
			\vertex [above =2.0cm of a2MDA](a2MDAAB) {\(\scalebox{1.85}{$H$}\)};
			\vertex [right =1.0cm of a2MDA](a2MDALab) {\(\scalebox{1.85}{$\ell$}\)};
			\vertex [left =1.0cm of a2MDA](a2MDALableft) {\(\scalebox{1.85}{$e$}\)};
			\diagram*{
			(a2BL)--[plain, thick](a2)--[double, thick, edge label=\(\scalebox{1.85}{$\varphi$}\)](a2M)--[double, thick, edge label=\(\scalebox{1.85}{$\varphi$}\)](a2MM)--[plain, thick](a2MMBR),
			(a2M)--[photon, thick](a2MDB),
			(a2)--[plain, thick, half left, looseness=1.7](a2MM),
			(a2MDA)--[dashed, thick](a2MDAAB)
			};
			\draw[fill=red] (a2M) circle(0mm);
			\end{feynman}
	\end{tikzpicture}
&
\begin{tikzpicture}[baseline=(a2.base), scale=0.54, transform shape]
			\begin{feynman}
			\vertex (a2);
			\vertex [above =2.0cm of a2](a2A);
			\vertex [below =2.0cm of a2](a2B);
			\vertex [left =2.0cm of a2A](a2AL) {\(\scalebox{1.85}{$e$}\)};
			\vertex [left =2.0cm of a2B](a2BL) {\(\scalebox{1.85}{$\ell$}\)};
			\vertex [right =2.0cm of a2](a2M) ;
			\vertex [right =2.0cm of a2M](a2MM);
			\vertex [above =2.0cm of a2MM](a2MMA);
			\vertex [below =2.0cm of a2MM](a2MMB);
			\vertex [right =2.0cm of a2MMA](a2MMAL) {\(\scalebox{1.85}{$\ell$}\)};
			\vertex [right =2.0cm of a2MMB](a2MMBL) {\(\scalebox{1.85}{$e$}\)};
			\vertex [below =2.0cm of a2M](a2MDB) ;
			\vertex [right =2.0cm of a2MM](a2MMright) ;
			\diagram*{
			(a2AL)--[plain, thick](a2)--[plain, thick](a2BL),
			(a2MM)--[plain, thick](a2MMAL),
			(a2MM)--[plain, thick](a2MMBL),
			(a2)--[double, thick, edge label=\(\scalebox{1.85}{$\varphi$}\)](a2MM)
			};
			\draw[fill=red] (a2M) circle(0mm);
			\end{feynman}
	\end{tikzpicture}
\\
\addlinespace[0.5em]
	\begin{tikzpicture}[baseline=(a2.base), scale=0.54, transform shape]
			\begin{feynman}
			\vertex (a3);
			\vertex [above =2.0cm of a3](a3A);
			\vertex [below =2.0cm of a3](a3B);
			\vertex [left =2.0cm of a3A](a3AL) {\(\scalebox{1.85}{$H$}\)};
			\vertex [left =2.0cm of a3B](a3BL) {\(\scalebox{1.85}{$\ell$}\)};
			\vertex [right =2.0cm of a3](a3M);
			\vertex [right =2.0cm of a3M](a3MM);
			\vertex [above =2.0cm of a3MM](a3MMA);
			\vertex [below =2.0cm of a3MM](a3MMB);
			\vertex [right =2.0cm of a3MMA](a3MMAL) {\(\scalebox{1.85}{$H$}\)};
			\vertex [right =2.0cm of a3MMB](a3MMBL) {\(\scalebox{1.85}{$e$}\)};
			\vertex [below =2.0cm of a3M](a3MDB) {\(\scalebox{1.85}{$H$}\)};
			\diagram*{
			(a3AL)--[dashed, thick](a3)--[plain, thick](a3BL),
			(a3)--[double, thick, edge label=\(\scalebox{1.85}{$F_1$}\)](a3M)--[double, thick, edge label=\(\scalebox{1.85}{$F_2$}\)](a3MM)--[dashed, thick](a3MMAL),
			(a3MM)--[plain, thick](a3MMBL),
			(a3M)--[dashed, thick](a3MDB),
			};
			\draw[fill=PadovaRed] (a3M) circle(1.38mm);
			\end{feynman}
	\end{tikzpicture}
    &
\begin{tikzpicture}[baseline=(a2.base), scale=0.54, transform shape]
			\begin{feynman}
			\vertex (a1);
			\vertex [below =2.0cm of a2](a2B) ;
			\vertex [left =2.0cm of a2B](a2BL) {\(\scalebox{1.85}{$\ell$}\)};
			\vertex [right =2.0cm of a2](a2M) ;
			\vertex [right =2.0cm of a2M](a2MM);
			\vertex [below =2.0cm of a2MM](a2MMB) ;
			\vertex [right =2.0cm of a2MMB](a2MMBR) {\(\scalebox{1.85}{$e$}\)};
			\vertex [below =2.0cm of a2M](a2MDB) {\(\scalebox{1.85}{$H$}\)};
			\vertex [above =2.0cm of a2M](a2MDA) ;
			\vertex [above =2.0cm of a2MDA](a2MDAAB) {\(\scalebox{1.85}{$B/W$}\)};
			\vertex [right =1.0cm of a2MDA](a2MDALab) {\(\scalebox{1.85}{$~H$}\)};
            \vertex [left =1.0cm of a2MDA](a2MDALableft) {\(\scalebox{1.85}{$H~$}\)};
			\diagram*{
			(a2BL)--[plain, thick](a2)--[double, thick, edge label=\(\scalebox{1.85}{$F_1$}\)](a2M)--[double, thick, edge label=\(\scalebox{1.85}{$F_2$}\)](a2MM)--[plain, thick](a2MMBR),
			(a2M)--[dashed, thick](a2MDB),
			(a2)--[dashed, thick, half left, looseness=1.7](a2MM),
			(a2MDA)--[photon, thick](a2MDAAB)
			};
			\draw[fill=PadovaRed] (a2M) circle(1.38mm);
			\end{feynman}
	\end{tikzpicture}
	&
	\begin{tikzpicture}[baseline=(a2.base), scale=0.54, transform shape]
			\begin{feynman}
			\vertex (a3);
			\vertex [above =2.0cm of a3](a3A);
			\vertex [below =2.0cm of a3](a3B);
			\vertex [left =2.0cm of a3A](a3AL) {\(\scalebox{1.85}{$H$}\)};
			\vertex [left =2.0cm of a3B](a3BL) {\(\scalebox{1.85}{$f$}\)};
			\vertex [right =2.0cm of a3](a3M);
			\vertex [right =2.0cm of a3M](a3MM);
			\vertex [above =2.0cm of a3MM](a3MMA);
			\vertex [below =2.0cm of a3MM](a3MMB);
			\vertex [right =2.0cm of a3MMA](a3MMAL) {\(\scalebox{1.85}{$H$}\)};
			\vertex [right =2.0cm of a3MMB](a3MMBL) {\(\scalebox{1.85}{$f$}\)};
			\vertex [below =2.0cm of a3M](a3MDB);
			\diagram*{
			(a3AL)--[dashed, thick](a3)--[plain, thick](a3BL),
			(a3)--[double, thick, edge label=\(\scalebox{1.85}{$F_{1,2}$}\)](a3MM),
			(a3MM)--[dashed, thick](a3MMAL),
			(a3MM)--[plain, thick](a3MMBL)
			};
			\draw[fill=red] (a3M) circle(0mm);
			\end{feynman}
	\end{tikzpicture}
    \end{tabular}
\caption{Representative diagrams for the single field configuration involving $\varphi$ (upper row), and a two-field configuration featuring two fermionic mediators (lower row) generating the operators discussed in Sec.~\ref{sec:Matching}. The portal interaction is indicated by the red dot, while $f$ denotes either $\ell$ or $e$. See sections ~\ref{sec:single_field_extensions} and~\ref{sec:two_field_extensions} for the matching expressions.} 
\label{fig:diagrams_combined}
\end{figure*}

Finally, the $[\cO_{\ell e}]_{ijk\ell}=(\bar\ell_i\gamma^\mu \ell_j)(\bar e_k \gamma_\mu e_\ell)$ four-fermion SMEFT operator generated by $\varphi$ at tree level is phenomenologically important, with the Wilson coefficient given by
\begin{equation}\label{eq:Ole_varphi}
    [\cC_{\ell e}]_{ijk\ell}=-\frac{1}{2M_\varphi^2}[Y_\varphi^e]_{i \ell}[Y_\varphi^e]^*_{jk}\,.
\end{equation}
We show representative diagrams involving $\varphi$ that generate the operators discussed in this section in Fig.~\ref{fig:diagrams_combined} (upper row).

Beyond the effects discussed above, $\varphi$ induces additional operators at the one-loop order. For instance, we explicitly checked the matching onto the $\cO_{HD}$ operator, which affects the oblique $T$-parameter. In this setup, however, the one-loop coefficient is proportional to $g_\sscript{Y}^2$, rendering the effect irrelevant for the phenomenological considerations presented below. Similarly, $\varphi$ generates one-loop corrections to Higgs-fermion current operators, with the least suppressed terms scaling as $g_{\sscript{Y},\sscript{L}}^2(Y_\varphi^e)^2$. The induced modifications to $Z$-lepton vertices are likewise loop-suppressed and phenomenologically irrelevant.

\subsubsection{Two-field extensions}
\label{sec:two_field_extensions}
Simplified extensions involving two new fields represent a richer class of UV completions for the $\mathcal{O}_{eH}$ operator at tree level, capable of inducing CP violating effects. In contrast to single-field extensions, where the structure of the resulting contributions is often too constrained to generate CP violation, the two-field configurations provide additional freedom through new portal interactions that enter directly in the matching relations. Depending on the spin and gauge representation of the mediators, the relevant configurations can involve either two scalar fields $ (S_1,S_2)$, two vector-like fermions (VLFs) $(F_1,F_2)$, or a scalar and a vector-like fermion $(S,F)$~\cite{deBlas:2017xtg}.\footnote{Although scalar-vector $(S,V)$ extensions are in principle possible, we do not include such configurations in our analysis. Part of the justification aligns with the discussion in Sec.~\ref{sec:single_field_extensions}, where vector interactions with the scalar sector are shown to yield real contributions, limiting their ability to induce CP violation.} All two-field configurations considered in this work are summarized in Tab.~\ref{tab:config_summary}, which lists the combinations of heavy fields along with the relevant UV interaction Lagrangians. 

In what follows, we provide a detailed analysis of the effective modifications of the Higgs–lepton Yukawa couplings, along with the associated contributions to lepton dipole moments, as well as four-fermion ($\psi^4$) and Higgs-current ($\psi^2 H^2 D$) operators relevant for all the constraints considered in this work. For illustrative purposes, in Fig.~\ref{fig:diagrams_combined} (lower row) we show some of the representative diagrams generating the operators discussed here in the scenario with two vector-like fermions $(F_1,F_2)$.

\vspace{0.2cm}
\noindent
{\textbf{Yukawa modifications.}} 
These modifications arise through the tree-level matching onto the $\cO_{eH}$ operator, and can lead to CP-odd contributions to the lepton Yukawas. The Wilson coefficient $\cC_{eH}$ resulting from two-field configurations can be expressed in terms of the UV couplings and mass parameters associated with each mediator pair, with the precise structure depending on the spin and gauge quantum numbers of the fields involved:
\begin{equation}\label{eq:CeH_two_field_config}
    [\cC_{eH}]_{ij} =
    \begin{cases}
        \begin{array}{@{}l@{\quad}l}
             \dfrac{\alpha(S_1,\varphi)}{M_{S_1}^2 M_\varphi^2} X(S_1,\varphi)[Y_\varphi^e]_{ij} & (S_1,\varphi)
            \\\\
             \dfrac{\alpha(F_1, F_2)}{M_{F_1} M_{F_2}} X_{ij}(F_1, F_2) & (F_1, F_2)~~~~~~~~~~~.
             \\\\
             \dfrac{\alpha(S, F)}{M_S^2 M_F} X_{ij}(S, F) & (S, F)
        \end{array}
    \end{cases}
\end{equation}
The numerical prefactors $\alpha$ and the UV coupling structures $X$ that enter these expressions are collected for all relevant two-field configurations in Tab.~\ref{tab:2field_matching_params}. In addition to the contributions included in Eq.~\eqref{eq:CeH_two_field_config}, two-field configurations also feature individual contributions to $\cC_{eH}$ arising from diagrams involving a single mediator field. However, these terms are strictly proportional to the SM Yukawa couplings as demonstrated in Eq.~\eqref{eq:CeH_Delta1}. As a result, these contributions are safely neglected in most of our analysis, with the exception of the final scenario considered in Sec.~\ref{sec:multiflavor}, in which the MFV alignment of both the individual and the portal couplings implies a similar scaling of the two types of contributions.

\vspace{0.2cm}
\noindent
{\textbf{Dipole operators.}} The contributions to lepton electric dipole moments in this framework arise at one loop through the generation of the SMEFT dipole operators, as discussed in Sec.~\ref{subsec:dipole_moms}. 
Among the two-field UV completions considered, the fermion-fermion $(F_1,F_2)$ and scalar-fermion $(S,F)$ ones yield genuinely new contributions to the dipole operators, while the $(S,S)$ configuration is discussed separately below. 
In the first two cases, the EDM for a generic lepton flavor can be parameterized with the LEFT dipole Wilson coefficient\footnote{Note that the one-loop matching is performed using \texttt{Matchete}~\cite{Fuentes-Martin:2022jrf} under the assumption that the two heavy mediators are approximately degenerate in mass. This justifies a common matching scale for both fields, which manifests as a unified effective mass scale in the denominators of Eq.~\eqref{eq:dipole_2_field_gen}. }
\begin{equation}\label{eq:dipole_2_field_gen}
    [\cC_{e\gamma}]_{ij}=\frac{ev \sqrt2}{32\pi^2}\times 
    \begin{cases}
        \begin{array}{@{}l@{\quad}l}
             \dfrac{\beta(F_1,F_2)}{M_{F_{1,2}}^2}X_{ij}(F_1,F_2) & (F_1,F_2)
            \\\\
             \dfrac{\beta(S,F)}{M_{F,S}^3}X_{ij}(S,F)  & (S,F)
        \end{array}
    \end{cases}\,.
\end{equation}
Here, $\beta(\mathcal{F}_1,\mathcal{F}_2)$ denotes the numerical factors, while the functions $X(\mathcal F_1,\mathcal F_2)$ encode the flavor structures constructed from the underlying UV couplings. Importantly, these flavor structures coincide with those appearing in Eq.~\eqref{eq:CeH_two_field_config}, underscoring the interplay between the two operators and giving rise to intrinsically correlated phenomenology. For each pair $(\mathcal F_1,\mathcal F_2)$, the explicit forms of $\beta$ and $X$ are collected in Tab.~\ref{tab:2field_matching_params}. Analogously to $\cC_{eH}$, one-loop diagrams involving individual mediator fields induce dipole operator contributions; however, because these are aligned with the SM Yukawas, they will be omitted from most of the following discussion and only considered when explicitly required.

Finally, in the case of two-scalar UV completions, as indicated in Tab.~\ref{tab:config_summary}, there exist three distinct configurations, each involving the $\varphi$ scalar. These setups feature a portal-type interaction between $\varphi$ and an additional scalar field, while only $\varphi$ couples directly to SM fermions via Yukawa-like interactions. Consequently, the leading contributions to the SMEFT dipole operators originate solely from the $\varphi$ scalar, with neither the second scalar nor the scalar portal playing an active role in the dipole matching. At the diagrammatic level, only the Yukawa coupling of $\varphi$ enters the one-loop amplitude relevant for the dipole operator (see Eq.~\eqref{eq:dipole_op_varphi}).

\vspace{0.2cm}
\noindent
$\bm{\psi^4}\textbf{ and }\bm{\psi^2H^2D.}$ In context of two-field extensions, SMEFT operators of the $\psi^4$ class are generated at tree level only in specific configurations involving scalar mediators. In particular, two-field configurations featuring $\varphi$ and $\Xi_1$, through their interactions with the lepton bilinears (see Tab.~\ref{tab:config_summary}), allow for the tree-level generation of $\cO_{\ell e}$ (with Wilson coefficient given by Eq.~\eqref{eq:Ole_varphi}) as well as the $[\cO_{\ell\ell}]_{ijk\ell}=(\bar\ell_i\gamma^\mu \ell_j)(\bar\ell_k\gamma_\mu\ell_\ell)$ operator, with the Wilson coefficient
\begin{equation}
\label{eq:Ole_Xi}
    [\cC_{\ell\ell}]_{ijk\ell}=\frac{1}{M^2_{\Xi_1}}[y_{\Xi_1}]_{ki}[y_{\Xi_1}]^*_{\ell j}\,.
\end{equation}
On the other hand, operators of the $\psi^2 H^2 D$ class arise from integrating out VLFs at tree level. Depending on the structure of the UV interactions, the relevant SMEFT operators include $[\cO_{H\ell}^{(1)}]_{ij}=(H^\dag i\overleftrightarrow {D}_\mu  H)(\bar\ell_i \gamma^\mu\ell_j)$, $[\cO_{H\ell}^{(3)}]_{ij}=(H^\dag i\overleftrightarrow {D}^I_\mu  H)(\bar\ell_i \gamma^\mu \sigma^I \ell_j)$ and $[\cO_{He}]_{ij}=(H^\dag i\overleftrightarrow {D}_\mu  H)(\bar e_i \gamma^\mu  e_j)$. Their corresponding Wilson coefficients can be written as~\cite{deBlas:2017xtg}
    \begin{equation}\label{eq:EDelta3_Higgs_fermion_currents}
    \begin{alignedat}{2}
        [\cC_{H\ell}^{(1)}]^{F}_{ij}&=-\frac{\delta_{F,E}}{4M_E^2}[\lambda_E]^*_i[\lambda_E]_j
        +\frac{3\,\delta_{F,\Sigma}}{16M^2_\Sigma}[\lambda_\Sigma]^*_i[\lambda_\Sigma]_j
        \\&~~~\,-\frac{3\,\delta_{F,\Sigma_1}}{16M^2_{\Sigma_1}}[\lambda_{\Sigma_1}]^*_i[\lambda_{\Sigma_1}]_j
        \,,
        \\
        [\cC_{H\ell}^{(3)}]^{F}_{ij}&=-\frac{\delta_{F,E}}{4M_E^2}[\lambda_E]^*_i[\lambda_E]_j
        +\frac{\delta_{F,\Sigma}}{16M^2_\Sigma}[\lambda_\Sigma]^*_i[\lambda_\Sigma]_j
        \\&~~~\,+\frac{\delta_{F,\Sigma_1}}{16M^2_{\Sigma_1}}[\lambda_{\Sigma_1}]^*_i[\lambda_{\Sigma_1}]_j
        \,,
        \\
        [\cC_{He}]^{F}_{ij}&=\frac{\delta_{F,\Delta_1}}{2M^2_{\Delta_1}}[\lambda_{\Delta_1}]^*_i[\lambda_{\Delta_1}]_j-\frac{\delta_{F,\Delta_3}}{2M^2_{\Delta_3}}[\lambda_{\Delta_3}]^*_i[\lambda_{\Delta_3}]_j\,,
    \end{alignedat}
\end{equation}
where the Wilson coefficients are presented in a unified form, with the contributions from different VLFs distinguished using Kronecker deltas. For any given two-field configuration the expressions reduce to the corresponding contributions by selecting the appropriate VLF representations $F=\{E,\Sigma,\Sigma_1,\Delta_1,\Delta_3\}$ (see Tab.~\ref{tab:config_summary} for the overview of the two-field configurations featuring VLFs and the relevant UV interaction structures).

\subsection{Complementary constraints}
\label{sec:complementary_constraints}
As emphasized in Sec.~\ref{sec:Matching}, the simplified UV models capable of generating CP-violating $\mc{O}_{eH}$ at tree level also generate a plethora of additional operators, which in turn give rise to a variety of complementary constraints, in addition to the ones discussed in the EFT setting in Sec.~\ref{sec:EFT_pheno}. In what follows, we comment on these constraints in the context of the operators discussed in the previous section.

\vspace{0.2cm}
\noindent
{\textbf{EDMs. }}As indicated in Sec.~\ref{subsec:dipole_moms}, the $\cO_{eH}$ operator, generated by both single- and two-field configurations, directly feeds into two-loop Barr–Zee-type contributions to EDMs. However, a crucial feature of these UV completions is that they can also generate the dipole operator $\cO_{e\gamma}$ at the one-loop level, as given by Eqs.~\eqref{eq:dipole_op_varphi} and \eqref{eq:dipole_2_field_gen}. This leads to an additional and potentially leading contribution to the dipole moments of leptons.

\vspace{0.2cm}
\noindent
{\textbf{cLFV. }} Depending on the underlying flavor structure of the UV couplings, the simplified extensions considered here can generate effective operators that mediate cLFV transitions. In particular, dipole operators (see Eqs.~\eqref{eq:dipole_op_varphi} and \eqref{eq:dipole_2_field_gen}) contribute directly to radiative decays such as $\ell_i \to \ell_j \gamma$, while SMEFT operators from the $\psi^4$ (Eqs.~\eqref{eq:Ole_varphi} and \eqref{eq:Ole_Xi}) and $\psi^2H^2D$ (Eq.~\eqref{eq:EDelta3_Higgs_fermion_currents}) class induce a broader set of cLFV observables. These include three-body decays such as $\ell_i \to \ell_j\ell_j\bar\ell_j$ and $\ell_i \to \ell_j \ell_k \bar\ell_k$, as well as coherent $\mu \to e$ conversion in nuclei~\cite{Calibbi:2021pyh,Davidson:2020hkf,Ardu:2024bua,Crivellin:2013hpa,Pruna:2014asa,Uesaka:2024tfn}. The cLFV observables considered in this work are collected in App.~\ref{app:cLFV_EDelta3}. In principle, also lepton flavor violating $\cC_{eH}$ contributes to cLFV processes, mediated by the Higgs boson~\cite{Harnik:2012pb, Dorsner:2015mja}. We find that such effects are typically suppressed by powers of the leptonic Yukawa couplings relative to the leading direct UV contributions.

\vspace{0.2cm}
\noindent
{\textbf{EWPT/LEP. }}Precision measurements of electroweak observables provide important constraints on the considered scenarios. The $\psi^2 H^2 D$ operators (Eq.~\eqref{eq:EDelta3_Higgs_fermion_currents}), generated by VLFs at tree level, modify the couplings of charged leptons to the electroweak gauge bosons. These vertex corrections are probed by $Z$-pole observables, such as partial decay widths, forward–backward asymmetries, and lepton universality tests~\cite{ALEPH:2005ab, Efrati:2015eaa, SLD:2000jop, Janot:2019oyi, ALEPH:2005ab, ParticleDataGroup:2024cfk} (see e.g.~\cite{Efrati:2015eaa} for a list of observables). In the following phenomenological analysis, we use the EWPT likelihood implemented in \texttt{smelli}~\cite{Aebischer:2018iyb,Stangl:2020lbh, Straub:2018kue, Aebischer:2018bkb, Freitas:2014hra, Brivio:2017vri}. In addition, scalar-mediated scenarios can generate four-lepton contact operators (Eqs.~\eqref{eq:Ole_varphi} and \eqref{eq:Ole_Xi}), which contribute to $e^+e^-\to \ell^+\ell^-$ scattering, modifying the cross sections and forward-backward asymmetries measured at LEP-II at energies above the $Z$ pole~\cite{ALEPH:2013dgf}. We use the likelihood presented in Refs.~\cite{Falkowski:2015krw,Falkowski:2017pss} for the subsequent phenomenological analysis, where such LEP-II constraints are phenomenologically important. For completeness, we have also considered the remaining low-energy constraints included in Refs.~\cite{Falkowski:2015krw,Falkowski:2017pss}. While these observables can in principle yield nontrivial sensitivity, their numerical impact is consistently subleading relative to the constraints emphasized in this work. Finally, we have also checked that the $O_{HD}$ operator, generated at one loop in all scenarios, as well as the $\psi^2 H^2 D$ generated at the one-loop level in the $\varphi$ scenarios, lead only to subleading bounds compared to the ones presented in this work.

\vspace{0.2cm}
\noindent
{\textbf{Direct searches. }} Searches for heavy vector-like leptons (VLLs) at colliders provide another class of constraints on the simplified UV completions considered here. At the LHC, both single- and pair-production channels have been explored, leading to lower bounds on the masses of charged VLLs typically in the few-hundred GeV range, depending on their electroweak quantum numbers and decay modes~\cite{ATLAS:2024mrr,CMS:2024bni,CMS:2022cpe,CMS:2025urb,ATLAS:2025wgc,ATLAS:2023sbu,CMS:2019hsm}. These searches are representative examples of the collider reach, but they do not exhaust the full set of analyses available. Importantly, the resulting limits remain far weaker than those obtained from low-energy precision probes such as EDMs or cLFV observables, and therefore do not play a competitive role in constraining the parameter space relevant for our study. Instead, they serve mainly to delineate the region of applicability of the EFT approach by ensuring that the masses of the VLLs are sufficiently above the electroweak scale. Finally, direct searches of heavy scalars coupling exclusively to leptons are challenging at the LHC, due to a highly suppressed production mechanism~\cite{deGouvea:2019qaz,BaBar:2020jma,Chen:2021rnl,Chen:2019hey,Belle:2022gbl,Afik:2023vyl}. We expect the rest of the observables considered in this work to be the dominant constraints on such scenarios.

\subsection{Flavor assumptions and phenomenology}
\label{sec:pheno}
The matching expressions derived in Sec.~\ref{sec:Matching} depend on the flavor structure of the underlying UV interactions, and the phenomenological implications of the resulting Wilson coefficients can differ substantially depending on the flavor assumptions made on the underlying UV theory. 
In this section, we study in detail two representative UV scenarios: the single-mediator case $\varphi\sim(\rep1,\rep2)_{1/2}$, discussed in Sec.~\ref{sec:single_field_extensions}, and the two-mediator case involving $E\sim(\rep1,\rep1)_{-1}$ and $\Delta_3\sim(\rep1,\rep2)_{-3/2}$, discussed in Sec.~\ref{sec:two_field_extensions}.\footnote{We select the $(E,\Delta_3)$ configuration as a representative example, since it captures the full set of relevant phenomenological effects. By contrast, a two-scalar configuration does not generate additional dipole operators through the portal interaction, whereas the scalar–fermion combinations yield both operator classes leading to phenomenology qualitatively similar to the $(E,\Delta_3)$ case, with the differences confined to the numerical factors listed in Tab.~\ref{tab:2field_matching_params}. Finally, the $\Xi_1$ and $\Sigma$ mediators are distinguished by their ability to realize the seesaw mechanism and generate the Weinberg operator; however, such scenarios are typically associated with higher NP scales and are not included among our explicit examples. Notably, the $(\Xi_1,\Sigma)$ configuration is characterized by an exact cancellation of the two-field dipole contribution. See Sec.~\ref{sec:two_field_extensions} for further details on the complementary operators generated for different configurations.} For each scenario, we explore the phenomenological consequences of several flavor assumptions. These range from setups in which only a single lepton flavor is active, to more general cases in which multiple flavors contribute simultaneously.

\begin{figure*}[t]
    \centering  
    \includegraphics[width=0.75\paperwidth]{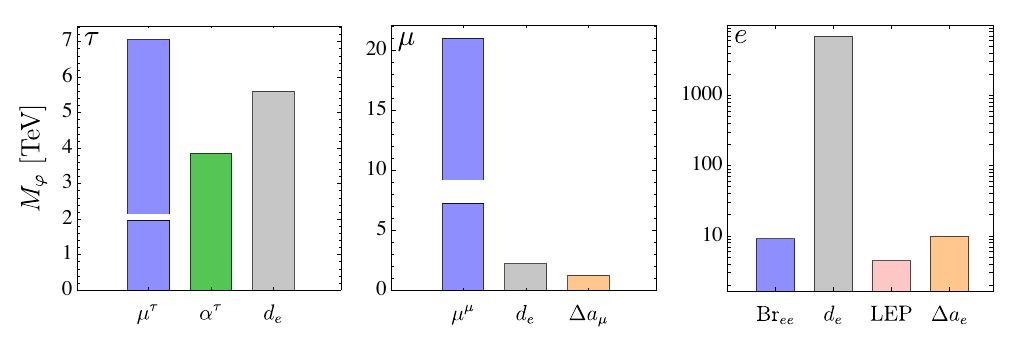}\\
    \includegraphics[width=0.75\paperwidth]{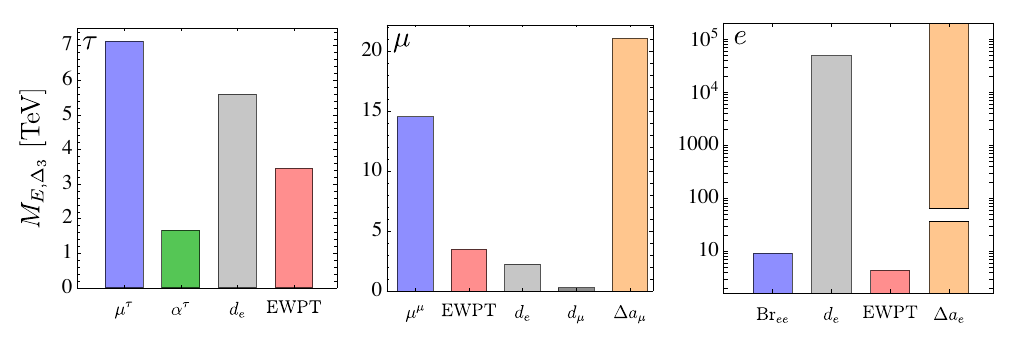}
    \caption{The 95\% CL bounds on the mass $M_\varphi$ of the scalar mediator $\varphi$ (upper row) and on the common mass scale $M_{E,\Delta_3}\equiv M_E=M_{\Delta_3}$ in the $(E, \Delta_3)$ two-field scenario (lower row). The three columns correspond to scenarios where the mediators couple exclusively to the $\tau$, $\mu$ and $e$ leptons, respectively. See Sec.~\ref{sec:single_flavor_assumptions} for a discussion.}
    \label{fig:single_bar_charts}
\end{figure*}

\subsubsection{Single-flavor assumptions}
\label{sec:single_flavor_assumptions}
Here we consider scenarios in which the new physics couplings involve only one lepton flavor at a time. Specifically, this implies that the independent flavor tensors entering the UV interactions (see e.g.~Tab.~\ref{tab:config_summary}) are restricted such that the same flavor index appears at each vertex. In each scenario, we adopt a benchmark choice where the representative couplings are set to unity (see discussion below), and extract the bounds on the (degenerate) masses of the UV degrees of freedom at the 95\% CL.

\vspace{0.2cm}
\noindent
\textbf{Scenario with $\bm{\varphi}$.} The tree-level matching onto $\cO_{eH}$ is provided in Eq.~\eqref{eq:varphiCeH}. For the numerical analysis we set $\lambda_\varphi=1$ and consider diagonal Yukawa entries of the form $[Y_\varphi^e]_{jj}=(1+i)/\sqrt{2}$ for $j={1,2,3}$, corresponding to $e$-, $\mu$-, and $\tau$-specific benchmark scenarios, respectively. For each case, we derive the bounds on the mediator mass $M_\varphi$ by considering the relevant experimental constraints one by one. The resulting constraints are presented in the upper row of Fig.~\ref{fig:single_bar_charts}. 

In case $\varphi$ couples exclusively to $\tau$ (Fig.~\ref{fig:single_bar_charts}, upper left), there are only three phenomenologically relevant constraints: the $h\to\tau^+\tau^-$ observables $\mu^\tau$ and $\alpha^\tau$, discussed in Sec.~\ref{subsec:direct}, and the electron EDM through the Barr-Zee diagrams, discussed in Sec.~\ref{subsec:dipole_moms}. In this selected direction in the parameter space (diagonally on Fig.~\ref{fig:CeH}, left), $\mu^\tau$ offers the best sensitivity to $M_\varphi$, with the reach of about $7\,\tev$. Note that there is a small allowed mass region around $2\,\tev$ due to the second crossing of the diagonal line in Fig.~\ref{fig:CeH} with the $\mu^\tau$ contour. This solution is challenged by both complementary constraints. The angular observable $\alpha^\tau$ offers a sensitivity of about $4\,\tev$, while electron EDM reaches about $5.5\,\tev$. There are no other noteworthy constraints on this scenario.

When $\varphi$ couples exclusively to $\mu$ (Fig.~\ref{fig:single_bar_charts}, upper middle), the dominant constraint comes from the collider measurements of $\mu^\mu$, in line with the results in Sec.~\ref{sec:EFT_results}, reaching a sensitivity of about $20\,\tev$, with a second solution possible around $8\,\tev$. As discussed in Sec.~\ref{sec:EFT_results}, the electron EDM constraint via the Barr-Zee effect is subdominant, offering sensitivity of only about $2\,\tev$. We additionally show the constraint from the anomalous magnetic moment of the muon which is generated in two ways: either directly via the one-loop contribution in Eq.~\eqref{eq:dipole_op_varphi}, or through the Barr-Zee effect discussed in Sec.~\ref{subsec:dipole_moms}. The second effect, leading to a constraint of about $1.5\,\tev$, is more important, as the direct one-loop contribution is Yukawa suppressed.

Lastly, in the case where $\varphi$ couples exclusively to electrons (Fig.~\ref{fig:single_bar_charts}, upper right), the sensitivity is completely dominated by the electron EDM constraint via the Barr-Zee effect, setting a limit of about $7~\mathrm{PeV}$. Note that $\varphi$ could in principle generate eEDM at one loop via Eq.~\eqref{eq:dipole_op_varphi}, however in this scenario $\cC_{e\gamma}$ is real. The collider constraints on $h\to e^+ e^-$ are nevertheless impressively sensitive to mass scales of about $10\,\tev$. A novel constraint due to the UV completion comes from LEP-II measurements of $e^+e^-\to e^+e^-$ scattering, as discussed in Sec.~\ref{sec:complementary_constraints}, which is sensitive to about $4.5\,\tev$. Finally, $\Delta a_e$ is generated in this setting, predominantly via the Barr-Zee effect as the direct one-loop contributions are Yukawa suppressed. The chosen benchmark point happens to go in the opposite direction to what is necessary for explaining the tension in $\Delta a_e$ discussed in Sec.~\ref{sec:dipole_exp_status}, and we obtain a lower limit on the mass of about $10\,\tev$.

\vspace{0.2cm}
\noindent
\textbf{Scenario with $\bm{(E,\Delta_3)}$.} Following a similar approach to the previous example, we consider $e$-, $\mu$-, and $\tau$-specific benchmark scenarios separately by setting $\lambda_{E\Delta_3}=(1+i)/\sqrt{2}$ and $[\lambda_E]_j = [\lambda_{\Delta_3}]_j = 1$ for $j={1,2,3}$ individually, see Tab.~\ref{tab:2field_matching_params}. Note that to ensure a CP-violating contribution in this setup, we introduce the complex phase through the portal interaction $\lambda_{E\Delta_3}$. The resulting constraints on the mass scale are summarized in the lower row of Fig.~\ref{fig:single_bar_charts}. 

In case $(E,\Delta_3)$ couple exclusively to $\tau$ (Fig.~\ref{fig:single_bar_charts}, lower left), the situation is very similar to the situation of $\varphi$, with some notable differences. First, by comparing Eqs.~\eqref{eq:varphiCeH} and \eqref{eq:CeH_two_field_config} (with negative $\beta$), the two coefficients differ in sign. Hence we are taking the opposite direction in the $\cC_{eH}$ plane in Fig.~\ref{fig:CeH} (left). This means that only a lower limit is obtained from $\mu^\tau$, again of about $7\,\tev$. The $\alpha^\tau$ constraint, due to its asymmetric nature, leads to a substantially different constraint of about $1.5\,\tev$. Electron EDM, through the Barr-Zee effect, reaches a similar sensitivity of $5.5\,\tev$ as before. Finally, there is an additional constraint purely from the UV dynamics due to electroweak precision tests, as discussed in Sec.~\ref{sec:complementary_constraints}. This set of observables reaches a constraint of about $3.5\,\tev$, and is dominated by measurements of $Z\to \tau^+\tau^-$.

When $(E,\Delta_3)$ couple exclusively to $\mu$ (Fig.~\ref{fig:single_bar_charts}, lower middle), the situation is drastically different compared to the case of $\varphi$. This is due to the anomalous magnetic moment of the muon being generated at the one-loop level with no additional suppression via Eq.~\eqref{eq:dipole_2_field_gen}, reaching a constraint of just above $20\,\tev$. The collider constraint from $\mu^\mu$ is slightly weakened compared to the previous scenario, again because of a different direction taken in the $\cC_{eH}$ plane, now resulting in the limit of about $15\,\tev$. The electron EDM constraint is due to Barr-Zee contributions, leading to a constraint of about $2\,\tev$. The EWPT sensitivity of about $3.5\,\tev$ is due to two effects: $Z\to \mu^+ \mu^-$ measurements, and the redefinition of $G_F$ (see e.g.~\cite{Brivio:2017vri}). Finally, we show the muon electric dipole moment constraint, primarily generate via direct one-loop contributions, which offers very poor sensitivity to masses of about $300\,\gev$.

Finally, when $(E,\Delta_3)$ couple exclusively to electrons (Fig.~\ref{fig:single_bar_charts}, lower right), the situation is similar to the case with $\varphi$. Again the leading constraint is by far due to electron EDM, this time reaching a staggering $50\,\mathrm{PeV}$ due to the unsuppressed one-loop UV contributions. The sensitivity from collider data on $h\to e^+e^-$ is comparable to that obtained previously, reaching around $10\,\tev$. The anomalous magnetic moment of the electron $\Delta a_e$ is sensitive to higher mass scales compared to the $\varphi$ case, once again due to the unsuppressed one-loop contributions. The benchmark point in this case happens to go in the correct direction to explain the tension in $\Delta a_e$ discussed in Sec.~\ref{sec:dipole_exp_status}, hence a range of allowed masses at the 95\% CL is shown, from about $35\,\tev$ to $65\,\tev$ (the SM is excluded at this level). In absence of any tension, the expected sensitivity of $\Delta a_e$ would be of this order of magnitude. EWPT constraints are sensitive to about $4.5\,\tev$ due to similar effects as already mentioned in the $\mu$-specific case.

\subsubsection{Multi-flavor assumptions}
\label{sec:multiflavor}
While the single-flavor assumptions are useful for comparing the individual lepton phenomenology in the UV context, we now turn to more realistic flavor assumptions, in which the UV degrees of freedom couple to more than one lepton flavor in various ways. As we demonstrate below, such scenarios can lead to interesting new complementary constraints, in particular due to charged lepton flavor violation (cLFV).

\begin{figure*}[t]
    \centering  \includegraphics[width=1.0\linewidth]{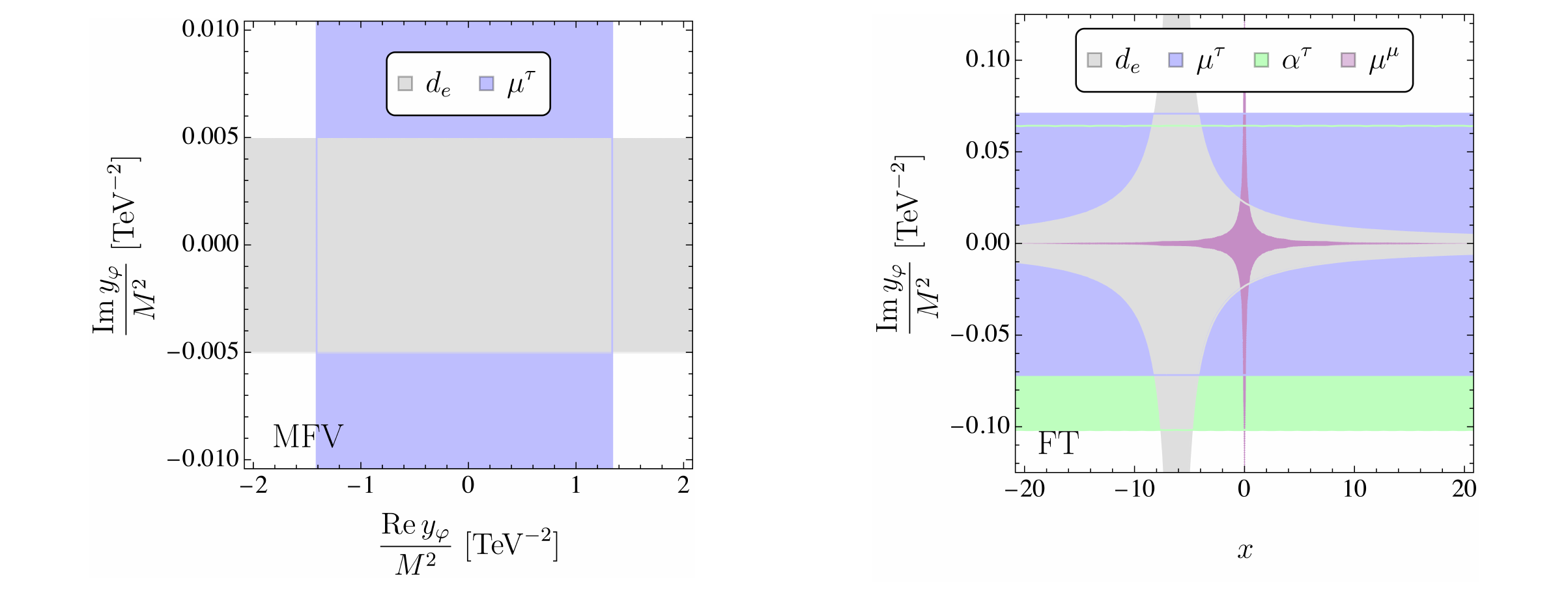}
    \caption{
    The 95\% CL bounds on the parameters of the $\varphi$ scenario under two flavor assumptions: MFV (left plot) and a fine tuned (FT) scenario with a vanishing electron coupling, while the muon and tau couplings enter proportionally with a relative weight $x$ (right plot). See Sec.~\ref{sec:multiflavor} for a discussion.}
    \label{fig:phi}
\end{figure*}

\vspace{0.2cm}
\noindent
\textbf{Scenario with $\bm{\varphi}$.} As indicated in Eq.~\eqref{eq:varphiint}, the flavor of the UV is completely determined by the $Y_\varphi^e$ coupling matrix. While various flavor assumptions could be imposed on this coupling matrix, we adopt a natural choice that minimizes the number of free UV parameters, namely, the framework of Minimal Flavor Violation (MFV)~\cite{DAmbrosio:2002vsn}. Given that the SM leptons transform under their respective $U(3)$ flavor groups, the SM Lagrangian augmented by UV-mediated interactions can be rendered formally invariant under $U(3)_\ell\times U(3)_e$ by treating both Higgs Yukawa $Y_\sscript{H}^e$ and $Y_\varphi^e$ as spurions transforming as $(\rep3_\ell,\repbar{3}_e)$. We therefore assume that both terms couple to the same flavon field and become proportional upon flavor symmetry breaking:
\begin{equation}\label{eq:varphi_MFV}
    [Y_\varphi^e]_{ij} = y_\varphi [Y^e_\sscript{H}]_{ij}\,, 
    \qquad
    y_\varphi \in \mathbb{C}\,.
\end{equation}
In this setting, the CP structure of the modified Higgs Yukawa couplings after electroweak symmetry breaking is governed by two complex UV parameters: $y_\varphi$ and $\lambda_\varphi$. For simplification, in the remainder of the analysis we assume $\lambda_\varphi=1$, and plot the constraints in the $(\re\,y_\varphi/M_\varphi,\, \im \,y_\varphi/M_\varphi)$ plane. The 95\% CL constraints on this scenario are presented in the left panel of Fig.~\ref{fig:phi}. The imaginary part of $y_\varphi$ is stringently constrained by the electron EDM. It is interesting to note that in this case, the dominant contribution is due to the Barr-Zee contribution with the electron coefficient $[\cC_{eH}]_{11}$ (see Eq.~\eqref{eq:BarrZee_semi_numerical}), despite the MFV suppression. As can be seen from Eq.~\eqref{eq:dipole_op_varphi}, the direct one-loop contributions are real in this scenario and therefore do not induce an electron EDM. The real part of $y_\varphi$ is best constrained from the $h\to \tau^+ \tau^-$ signal strength measurements discussed in Sec.~\ref{subsec:direct} due to i) the MFV alignment and ii) the most precise determination of a signal strength. Although the $\varphi$ scalar generates the $\cO_{\ell e}$ SMEFT operator at tree level (see Eq.~\eqref{eq:Ole_varphi}), the corresponding Wilson coefficient in the MFV scenario is proportional to $(Y_\sscript{H}^e)^\dag Y_\sscript{H}^e\sim y_\ell^2$, rendering its contribution to $e^+e^- \to \ell^+\ell^-$ negligible. Likewise, the contributions to $\Delta a_e$ and $\Delta a_\mu$ are suppressed as indicated in Eq.~\eqref{eq:dipole_op_varphi}, whereas the Barr–Zee contributions to these observables yield only subleading constraints. Lastly, the MFV structure ensures that no charged lepton flavor violation is induced, as the relevant operators remain diagonal in flavor space.

We next consider an illustrative flavor assumption in which $\varphi$ couples to the second and third lepton generations, whereas its coupling to electrons is assumed to be negligible. In particular, we take
\begin{equation}
    [Y_\varphi^e]_{22} = x y_\varphi\,,\qquad [Y_\varphi^e]_{33} = y_\varphi\,,
\end{equation}
where $x$ parameterizes the ratio between the muon and the tau couplings, while all other entries of the Yukawa matrix are set to zero. Next, we assume that $y_\varphi$ is imaginary and $x$ is real, and we plot the constraints on this scenario in the $(x, \im \,y_\varphi)$ plane. The 95\% CL constraints for this flavor assumption are shown in the right panel of Fig.~\ref{fig:phi}. The gray contour shows the constraint due to the Barr-Zee contributions to the electron EDM. Given the flavor structure with muon and tau couplings entering proportionally with relative weight $x$, the Barr–Zee contributions can cancel between the $\mu$ and $\tau$ loops. As a result, the electron EDM constraint loses sensitivity at a specific value of $x\approx-6$. From the collider perspective, the $h\to \tau^+ \tau^-$ measurements of $\mu^\tau$ and $\alpha^\tau$ are sensitive to $\im\,y_\varphi$ and can close this flat direction. Note also that the constraint from $\alpha^\tau$ is asymmetric around the imaginary axis, reflecting the CP-sensitive nature of this observable, as already discussed in Sec.~\ref{sec:EFT_results}. However, the signal strength of $h\to \mu^+ \mu^-$ offers a better sensitivity in this case, especially for large contributions to $[\cC_{eH}]_{22}$ (large $|x|$), such as those needed to achieve the cancellation in the Barr-Zee contributions to eEDM. We have verified that all other observables considered in this work lead to subleading constraints on this scenario.

\begin{figure*}[t]
    \centering  \includegraphics[width=0.95\linewidth]{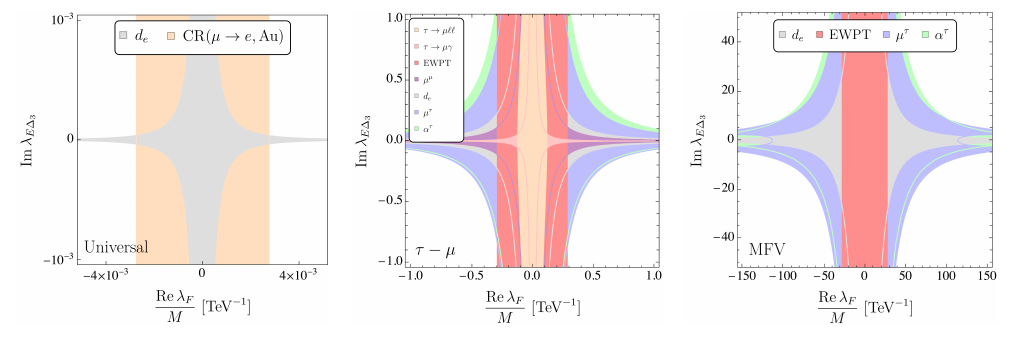}
    \caption{
    The 95\% CL bounds on the parameters of the $(E, \Delta_3)$ scenario under three flavor assumptions: universal (left plot), $\tau$ and $\mu$ specific (middle plot) and MFV (right plot). See Sec.~\ref{sec:multiflavor} for a discussion.
    }
    \label{fig:ED}
\end{figure*}

\vspace{0.2cm}
\noindent
\textbf{Scenario with $\bm{(E,\Delta_3)}$.} As indicated in Tab.~\ref{tab:config_summary}, this scenario features two independent flavor structures, encoded in the coupling vectors $\lambda_E$ and $\lambda_{\Delta_3}$. Once again, various flavor assumptions can be considered in this context, however we again aim to minimize the number of free parameters, while retaining phenomenologically interesting features. 

We first assume universal and equal couplings of both VLFs to the respective SM leptons:\footnote{Related coupling structures can arise in models with discrete flavor symmetries. For example, in $A_4$-based setups where both the SM leptons and the VLFs transform as triplets, symmetry dictates that only a single coupling parameter appears in the interaction term. While the resulting structure is diagonal in flavor space, it enforces equal couplings across generations, effectively yielding a universal interaction strength. More restricted flavor tensors, such as the $\mu-\tau$ structure considered below, can analogously be realized in setups based on flavor groups like $S_4$ that admit doublet representations~\cite{Palavric:2024gvu,Moreno-Sanchez:2025bzz}.}
\begin{equation}\label{eq:EDelta_3_scenario1}
    [\lambda_E]_i =[\lambda_{\Delta_3}]_i = \lambda_F\,, 
\end{equation}
In addition to this complex parameter, the UV interactions also contain the complex portal coupling $\lambda_{E\Delta_3}$. Upon matching onto the $\mathcal{O}_{eH}$ operator (see.~Eq.~\eqref{eq:CeH_two_field_config} and Tab.~\ref{tab:2field_matching_params}), the combination of couplings that enters is given as $|\lambda_F|^2 \lambda_{E\Delta_3}$. In order to streamline the analysis and still demonstrate the impact of potential new sources of CP violation, we fix $\lambda_F$ to be real and take $\lambda_{E\Delta_3}$ to be imaginary. We present the resulting constraints in the $(\re\,\lambda_F/M, \im\,\lambda_{E\Delta_3})$ parameter plane, as shown in the left panel of Fig.~\ref{fig:ED}. The stringent electron EDM constraint is dominated by the direct one-loop contributions due to Eq.~\eqref{eq:dipole_2_field_gen}, with Barr-Zee contributions being negligible. With universal couplings to all lepton flavors, this setup induces a range of cLFV processes mediated by the LFV $\psi^2 H^2 D$ (Eq.~\eqref{eq:EDelta3_Higgs_fermion_currents}) and dipole (Eq.~\eqref{eq:dipole_2_field_gen}) operators. The expressions of various cLFV branching ratios in the $(E, \Delta_3)$ scenario are collected in App.~\ref{app:cLFV_EDelta3}. In particular, $\mu\to e$ conversion in gold nuclei sets the most stringent constraint on $\re\,\lambda_F/M$. This enhanced sensitivity in case of $\mu\to e$ conversion arises because the process receives contributions not only from the dipole operator but also from the tree-level contact terms of the $\psi^2 H^2 D$ class, which are generated by integrating out the individual mediators. By contrast, $\mu \to e\gamma$ is mediated exclusively by the dipole operator carrying an additional loop suppression. The remaining observables considered in this work yield significantly weaker bounds and are therefore not competitive in constraining the parameter space under this assumption.

Given the stringent constraints from the electron EDM and cLFV including the first two generations, we next turn to a flavor scenario in which the new states couple exclusively to muons and taus:
\begin{equation}\label{eq:EDelta_3_scenario2}
    [\lambda_E]_i =[\lambda_{\Delta_3}]_i = \lambda_F \lzm\delta_{i2}+\delta_{i3}\dzm\,,
\end{equation}
Similarly to the previous case, we consider $\lambda_F$ to be real and $\lambda_{E\Delta_3}$ to be imaginary, and show the constraints in the plane of $(\re \,\lambda_F/M, \im \,\lambda_{E\Delta_3})$. As shown in the middle panel of Fig.~\ref{fig:ED}, the $\tau-\mu$ benchmark scenario leads to a significantly richer pattern of constraints of similar importance compared to the universal case. The two strongest constraints on $\re \,\lambda_F/M$ are due to LFV $\tau\to\mu\ell\ell$ decays and EWPT. A set of constraints is sensitive to the product of the two couplings, among which the $\tau\to\mu\gamma$ radiative cLFV decay leads to the dominant constraint. Among the collider constraints, the $h\to \mu^+\mu^-$ signal strength offers the best sensitivity in this scenario, beating the better measured $h\to \tau^+ \tau^-$ data. This is because the Higgs signal strength measurements are intrinsically sensitive to Higgs Yukawas of the order of the SM Yukawas - an universal effect shifts the muon Yukawa more (in relative terms) compared to tau Yukawa. In this scenario, the electron EDM can not be generated directly in the UV model at one loop, and it arises due to the Barr-Zee effect. It is interesting to note that $d_e$ is of similar sensitivity (slightly worse) compared to the collider $\mu^\mu$ data. Note that the anomalous magnetic moment of the muon $a_\mu$ is not generated in this scenario as both $\cC_{e\gamma}$ (which could induce it at one-loop level) and $\cC_{eH}$ (which could induce it through the Barr-Zee effect) are imaginary.

As a final example, we embed the $(E, \Delta_3)$ scenario in an MFV framework by promoting the vector-like leptons to flavor triplets (see e.g.~\cite{Greljo:2023adz,Greljo:2023bdy}), such that the interaction terms take the form\footnote{Alternatively, we could promote the flavor vectors to spurions of $U(3)_\ell\times U(3)_e$ as $\lambda_E\sim(\repbar{3}_\ell,\rep1)$,  $\lambda_{\Delta_3}\sim(1,\repbar{3}_e)$, and forbid the SM Yukawa couplings. This scenario would attempt to explain the structure of the SM Yukawas, but would alone lead to a rank 1 matrix, only explaining one of the masses, and would hence need to be extended. We decide not to study this scenario in this work.}
\begin{equation}\label{eq:EDelta3_MFV}
    \begin{alignedat}{2}
        -\cL_{\sscript{UV}}&\supset
        [\Lambda_E]_{ij} \bar{E}_{i} H^\dagger \ell_j
        +[\Lambda_{\Delta_3}]_{ij} \bar{\Delta}_{3i} H^\dagger e_j
        \\&
        +[\Lambda_{E\Delta_3}]_{ij} \bar{E}_{i} \tilde{H}^\dagger \Delta_{3 j}+\hermc\,.
    \end{alignedat}
\end{equation}
Under the assumption that the VLFs transform as $E\sim\rep3_e$ and $\Delta_3\sim \rep3_\ell$ under $U(3)_\ell\times U(3)_e$ flavor symmetry, the $\Lambda$ coupling tensors in Eq.~\eqref{eq:EDelta3_MFV} are promoted to flavor spurions, which transform as $\Lambda_E \sim (\repbar{3}_\ell, \rep3_e)$, $\Lambda_{\Delta_3} \sim (\rep3_\ell, \repbar{3}_e)$ and $\Lambda_{E\Delta_3} \sim (\repbar{3}_\ell, \rep3_e)$. Upon flavor symmetry breaking, these spurions can be aligned with the SM lepton Yukawa matrices, allowing us to express their values in terms of $Y^e_\sscript{H}$:
\begin{equation}
    \Lambda_E = \lambda_E (Y^{e}_\sscript{H})^\dag\,,\quad
    \Lambda_{\Delta_3} = \lambda_{\Delta_3} Y^{e}_\sscript{H}\,,\quad
    \Lambda_{E\Delta_3} = \lambda_{E\Delta_3} (Y^{e}_\sscript{H})^\dag\,,
\end{equation}
where $\lambda_E,\lambda_{\Delta_3},\lambda_{E\Delta_3}\in\mathbb C$. While the setup generally involves three independent complex UV parameters, we adopt a simplifying assumption analogous to earlier scenarios by setting the $E$ and $\Delta_3$ couplings equal and real:
\begin{equation}
\lambda_E =\lambda_{\Delta_3} = \lambda_F\,,
\qquad
\lambda_F \in \mathbb{R}\,,
\end{equation}
while we take $\lambda_{E\Delta_3}$ to be imaginary. As indicated above, an important feature of this MFV scenario is that the new VLFs are charged under the flavor symmetry group. As a result, the matching relations for the Wilson coefficients given by Eqs.~\eqref{eq:CeH_two_field_config}--\eqref{eq:EDelta3_Higgs_fermion_currents} must be amended by summing over internal flavor indices in the relevant diagrams. The resulting expressions for the Wilson coefficients of interest then take the form
\begin{equation}\label{eq:EDelta3_MFV_WCs_combined}
    \begin{alignedat}{2}
        [\cC_{eH}]_{ij}&=\frac{1-i\,\im\,\lambda_{E\Delta_3}}{M^2}|\lambda_F|^2\sum_k y_k^3\delta_{ik}\delta_{jk}\,,
        \\
        [\cC_{H\ell}^{(1)}]_{ij}&=[\cC_{H\ell}^{(3)}]_{ij}=-\frac{|\lambda_F|^2}{4M^2}\sum_ k y_k^2\delta_{i k}\delta_{j k}\,,
        \\
        [\cC_{He}]_{ij}&=-\frac{|\lambda_F|^2}{2M^2}\sum_ k y_k^2\delta_{i k}\delta_{j k}\,,
        \\
        [\cC_{e\gamma}]_{ij}&=\frac{-2+5\,i\,\im\lambda_{E\Delta_3}}{64\pi^2}\frac{e\,v}{\sqrt2}\frac{|\lambda_F|^2}{M^2}   \sum_k y_k^3\delta_{i k}\delta_{j k}  \,,
    \end{alignedat}
\end{equation}
where the summation index $k$ runs over the three lepton flavors. The factor $y_k$ denotes the charged-lepton Yukawa couplings in the mass basis. Note that in this scenario we explicitly take into account the single-field contributions to $\cC_{eH}$ and $\cC_{e\gamma}$. These are typically suppressed by the Higgs Yukawas, however in this scenario, in which the portal coupling between the two VLFs is aligned with the Higgs Yukawa, the suppression is the same between the single- and two-field contributions.

The constraints on this scenario in the $(\re \,\lambda_F/M, \im \,\lambda_{E\Delta_3})$ plane are presented in the right plot of Fig.~\ref{fig:ED}. Due to the MFV nature, charged lepton flavor is conserved, hence cLFV constraints are absent. The $\re \,\lambda_F/M$ direction is most constrained by EWPT, while the product of the two couplings is predominantly constrained by the electron EDM. Interestingly, even though eEDM is generated at the one-loop level in this scenario, it is suppressed by $y_e^3$. Hence the eEDM constraint is dominated by the tau contributions in the Barr-Zee type diagrams. Among the collider constraints, the most sensitive ones in the MFV scenario are the $h\to \tau^+\tau^-$ measurements. Note that the contour of $\mu^\tau$ in the $\re \,\lambda_F/M$ direction is closed due to the single-field contributions (see Eq.~\eqref{eq:EDelta3_MFV_WCs_combined}).

\begin{table*}[t]

\centering
\scalebox{0.93}{
\begin{tabular}{c@{\hspace{0.8cm}}c@{\hspace{0.8cm}}c@{\hspace{0.8cm}}c@{\hspace{0.8cm}}c@{\hspace{0.8cm}}c@{\hspace{0.8cm}}c@{\hspace{0.8cm}}c@{\hspace{0.8cm}}c}
\toprule
\textbf{cLFV obs.}
&\textbf{Bound}
&\textbf{Ref.}
&\textbf{cLFV obs.}
&\textbf{Bound}
&\textbf{Ref.}
&\textbf{cLFV obs.}
&\textbf{Bound}
&\textbf{Ref.}
\\[1pt]
\midrule
$\mathrm{BR}(\mu\to e\gamma)$
&$3.1\times10^{\eminus13}$
&\cite{MEGII:2025gzr}
&$\mathrm{BR}(\mu\to ee\bar e)$
&$1.0\times10^{\eminus12}$
&\cite{SINDRUM:1987nra}
&$\mathrm{BR}(\tau\to e\mu\bar\mu)$
&$2.7\times10^{\eminus8}$
&\cite{Hayasaka:2010np}
\\[3.5pt]
$\mathrm{BR}(\tau\to \mu\gamma)$
&$4.2\times10^{\eminus8}$
&\cite{Belle:2021ysv}
&$\mathrm{BR}(\tau\to ee\bar e)$
&$2.7\times10^{\eminus8}$
&\cite{Hayasaka:2010np}
&$\mathrm{BR}(\tau\to \mu e\bar e)$
&$1.8\times10^{\eminus8}$
&\cite{Hayasaka:2010np}
\\[3.5pt]
$\mathrm{BR}(\tau\to e\gamma)$
&$3.3\times10^{\eminus8}$
&\cite{BaBar:2009hkt}
&$\mathrm{BR}(\tau\to \mu\mu\bar \mu)$
&$1.9\times10^{\eminus8}$
&\cite{Hayasaka:2010np}
&$\mathrm{CR}(\mu\to e,\mathrm{Au})$
&$7.0\times10^{\eminus13}$
&\cite{SINDRUMII:2006dvw}
\\[2.5pt]
\bottomrule
\end{tabular}
}	
\caption{Overview of current experimental bounds on cLFV observables relevant to the $(E, \Delta_3)$ scenario. All bounds are given at 90\% confidence level.}
\label{tab:app:cLFV_obs}
\end{table*}

\section{Conclusions}
\label{sec:conc}
The CP nature of the Higgs Yukawa couplings is an important open question in high energy physics. It can be most directly determined via direct measurements at colliders, especially when CP-sensitive angular observables are viable. In this work, we focused on leptonic Higgs Yukawa couplings, and systematically studied the consequences of the assumption that they could be modified by sufficiently heavy new degrees of freedom. The aim was two-fold: determining a minimal set of heavy UV degrees of freedom that can induce CP-odd effects in leptonic Yukawas, and studying their rich complementary phenomenology. 

First, we parameterized generic short-distance NP contributions to leptonic Yukawas using the SMEFT. This framework alone implies nontrivial additional effects, correlated with the modifications of the Yukawa couplings. In particular, we demonstrated important complementary constraints due to electric and magnetic dipole moments. We find that the CP-odd part of the $\tau$ Yukawa is primarily constrained from the electron EDM, while the angular analyses of $h\to\tau^+\tau^-$ represent an important direct determination. In case of the $\mu$ Yukawa, the collider measurements of $h\to\mu^+\mu^-$ are currently the dominant constraint, albeit incapable of determining its exact CP nature. Finally, the CP-even part of the electron Yukawa is most strongly constrained by direct searches of $h\to e^+ e^-$, while the CP-odd part is very tightly constrained from the electron EDM.

We then took an important step towards a more realistic UV setting by determining the complete set of heavy scalar, fermion, and vector degrees of freedom, capable of generating CP-odd modifications of the lepton Yukawas at the tree level. Remarkably, only one single-field extension of the SM can account for this, namely the scalar $\varphi$ carrying the same gauge quantum numbers as the Higgs. Considering two-field extensions of the SM opens up a richer set of possibilities due to the existence of additional portal couplings. Introducing new heavy degrees of freedom inevitably induces additional effects, manifesting as a correlated set of higher-dimensional operators. Importantly, we found that these include dipole operators already at the one-loop level, and Higgs-current operators generated at tree level, all of which contribute to precise complementary observables.

Finally, we performed an extensive phenomenological analysis of two representative scenarios: the single scalar extension $\varphi$ and the extension with two vector-like fermions $(E, \Delta_3)$. In each of the scenarios, we studied multiple interesting flavor assumptions, either coupling the new degrees of freedom exclusively to a single lepton flavor, to pairs of lepton flavors, universally to all flavors, or consistently with an MFV assumption. For $\varphi$ we find that the SMEFT picture is often sufficient, with direct measurements at colliders and the Barr-Zee contributions to electron EDM playing the dominant role. On the contrary, in the case of $(E, \Delta_3)$, we find a much richer set of phenomenologically important complementary constraints, absent in the simplified SMEFT picture. These include unsuppressed contributions to electron EDM at one loop, cLFV observables such as $\mu \to e$ conversion and $\tau \to \mu$ decays, as well as electroweak precision tests.

The direct measurements of the Higgs properties at colliders remain an important avenue even in light of the induced correlated effects. Nevertheless, the highly precise complementary low energy probes play an unavoidable role in constraining realistic UV scenarios. We demonstrated the importance of a multi-faceted approach to searching for NP in the context of heavy degrees of freedom, capable of inducing CP-odd leptonic Yukawas. 

Looking ahead, a natural continuation of our present study is to apply the same framework to the quark Yukawas. Compared to the leptonic case, the quark sector presents a substantially richer phenomenology: a broader range of UV completions, intricate correlations with flavor-changing transitions and electric dipole moments, and potentially stronger collider signatures. A thorough analysis of these aspects promises to further clarify the CP properties of the Higgs Yukawas as well as the role of heavy new degrees of freedom.

\begin{acknowledgments}
AS and NK acknowledge the financial support from the Slovenian Research and Innovation Agency (grants No.~J1-50219, N1-0407 and research core funding No.~P1-0035). The work of AP has received funding from the Swiss National Science Foundation (SNF) through the Eccellenza Professorial Fellowship ``Flavor Physics at the High Energy Frontier'', project number 186866.
\end{acknowledgments}

\appendix

\section{Overview of cLFV observables for the $(E,\Delta_3)$ scenario}
\label{app:cLFV_EDelta3}

In this appendix, we collect the expressions for the cLFV observables induced in the $(E,\Delta_3)$ scenario, under the two flavor assumptions specified in Eqs.~\eqref{eq:EDelta_3_scenario1} and \eqref{eq:EDelta_3_scenario2}. The current experimental limits on the relevant cLFV channels discussed below are summarized in Tab.~\ref{tab:app:cLFV_obs}.

Starting with the radiative decays, the branching ratio for $\ell_i\to\ell_j\gamma$ transition can be expressed as~\cite{Calibbi:2021pyh}
\begin{equation}
    \begin{alignedat}{2}
        \mathrm{BR}(\ell_i\to \ell_j\gamma)&=\frac{m_i^3 v^2 e^2}{4\pi\Gamma_i}\frac{\beta(E,\Delta_3)^2}{(16\pi^2)^2}\frac{\re^4\lambda_F}{M^4}\im^2\lambda_{E\Delta_3}\,,
    \end{alignedat}
\end{equation}
where the index $i$ denotes the flavor of the decaying lepton, while $j$ labels the flavor of the final-state lepton. In the case of the universal flavor assumption, the full set of radiative decays is generated, including $\mu \to e \gamma$, $\tau \to \mu \gamma$, and $\tau \to e \gamma$. However, under the $\mu-\tau$ flavor assumption, only the $\tau \to \mu \gamma$ decay is induced, while processes involving the electron are absent.

In addition to the radiative transitions, the same set of UV interactions also generates various three-body cLFV decays. These fall into three distinct categories, depending on the flavor structure of the final-state leptons. The first class consists of processes of the form $\ell_i \to \ell_j \ell_j \bar{\ell}_j$, such as $\mu \to e e \bar{e}$, $\tau \to \mu \mu \bar{\mu}$, and $\tau \to e e \bar{e}$, which are directly induced by the diagrams involving a single flavor-changing current. The second class involves final states of the form $\ell_i \to \ell_j \ell_k \bar{\ell}_k$, with representative examples including $\tau \to e \mu \bar{\mu}$ and $\tau \to \mu e \bar{e}$. Both of these classes of transitions are relevant under the flavor assumptions defined in Eqs.~\eqref{eq:EDelta_3_scenario1} and \eqref{eq:EDelta_3_scenario2}. 
The branching ratios for these three-body decay modes can be expressed as~\cite{Calibbi:2021pyh}
\begin{widetext}
    \begin{equation}\label{eq:app_3body_cLFV}
        \begin{alignedat}{2}
        \mathrm{BR}(\ell_i\to \ell_j \ell_j \bar{\ell}_j)&=\frac{2\,m_i^5}{(16\pi)^3\Gamma_i}\frac{\re^4\lambda_F}{M^4}\lzs 1-4s_\sscript{W}^2+8s_\sscript{W}^4+\frac{e^4\beta(E,\Delta_3)^2}{6\pi^4}\frac{v^2}{m_i^2}\lzm\log\frac{m_i^2}{m_j^2}-\frac{11}{4}\dzm\im^2\lambda_{E\Delta_3} \dzs\,,
        \\
        \mathrm{BR}(\ell_i \to \ell_{j} \ell_k \bar\ell_k)&=\frac{10\,m_i^5}{3(16\pi)^3\Gamma_i}\frac{\re^4\lambda_F}{M^4}\lzs 1-4s_\sscript{W}^2+\frac{28}{5}s_\sscript{W}^4+\frac{e^4\beta(E,\Delta_3)^2}{10\pi^4}\frac{v^2}{m_i^2}\lzm\log\frac{m_i^2}{m_j^2}-3\dzm\im^2\lambda_{E\Delta_3} \dzs\,,
        \end{alignedat}
    \end{equation}
\end{widetext}
where the indices $i$ and $j$ label the initial and final state leptons, respectively, while the index $k$ denotes the flavor of the additional lepton pair in the final state. The absence of explicit dependence on $k$ in the branching ratios arises from the flavor structure assumed in both scenarios, where the UV couplings are either universal or confined to the $\mu-\tau$ sector with identical strengths. 

Final cLFV probe relevant to our analysis is coherent $\mu\to e$ conversion in gold nuclei. The conversion rate can be written as~\cite{Calibbi:2021pyh}
\begin{equation}\small
    \begin{alignedat}{2}
        \mathrm{CR}(\mu\to e,\mathrm{Au})&=\frac{m_\mu^5}{\Gamma_{\sscript{Au}}}\frac{\re^4\lambda_F}{M^4}\Big[9.62\times10^{\eminus3}+0.29\,\im^2\lambda_{E\Delta_3} \Big]\,,
    \end{alignedat}
\end{equation}
where $\Gamma_{\sscript{Au}}=8.60\times 10^{\eminus18}\,\gev$ denotes the total muon capture rate in gold nuclei~\cite{Kitano:2002mt} and where we directly plug the numerical values for the overlap integrals. The expression is relevant only for the scenario in which couplings to both muons and electrons are present. 

\bibliographystyle{JHEP}
\bibliography{main}

\end{document}